 \documentclass[pmlr]{jmlr}

 \RequirePackage{graphicx}
 \begingroup
 \catcode`\_=12
 \gdef\FSlideRegional{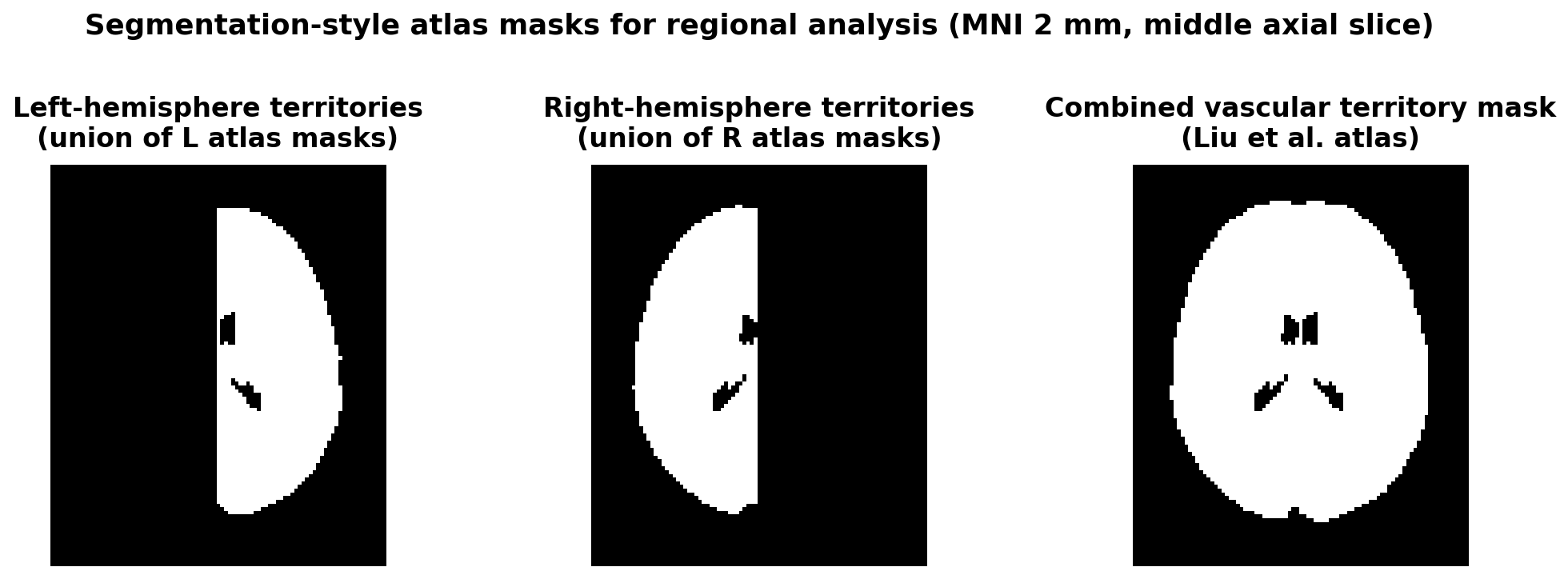}
 \gdef\FSlideVascular{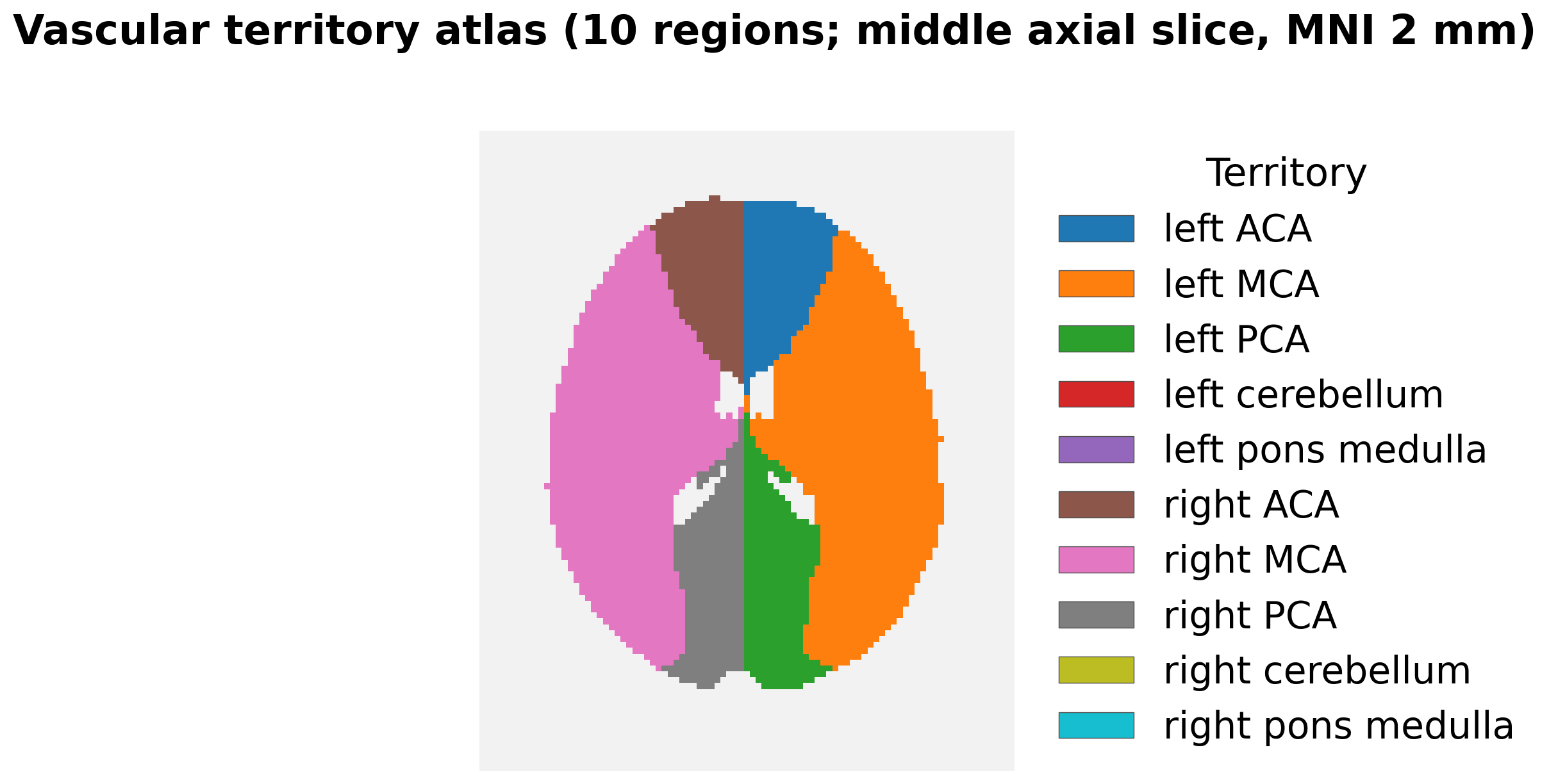}
 
 \gdef\FSlideGtDelta{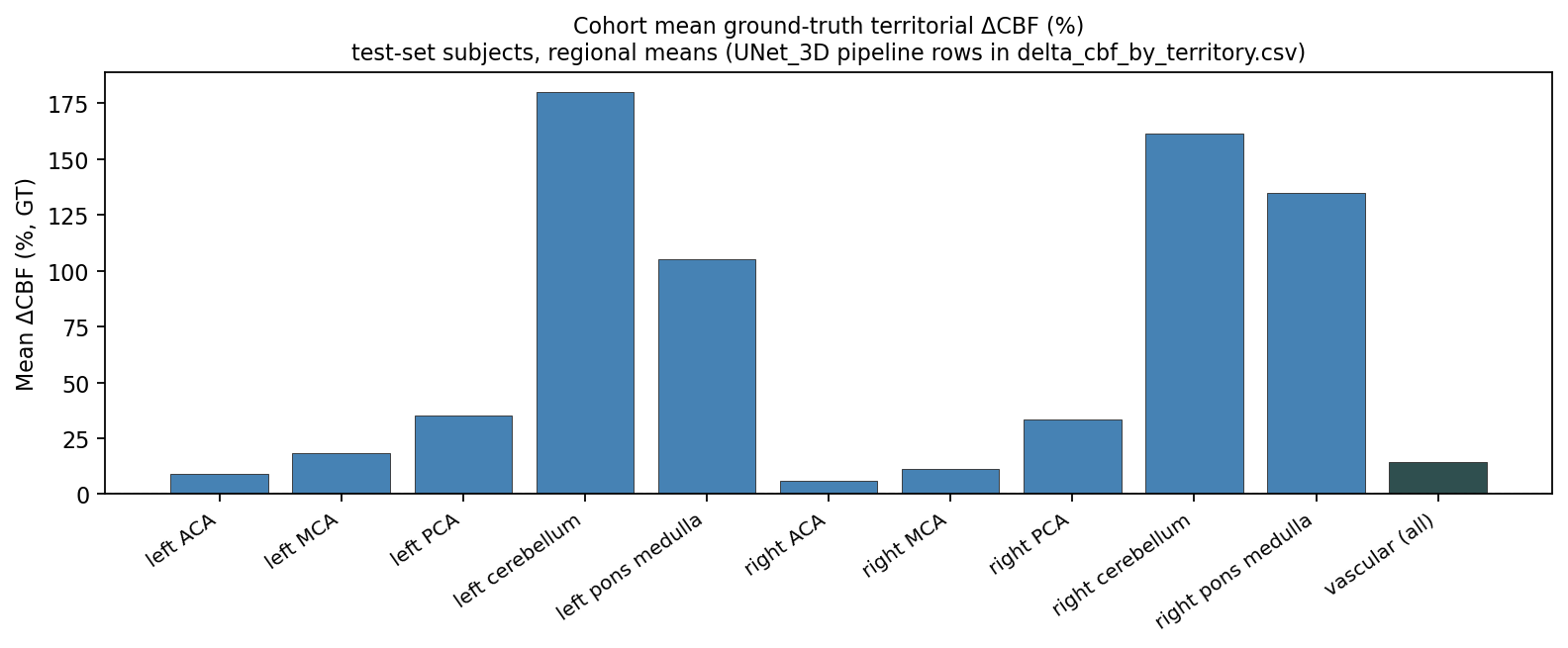}
 \gdef\FSlidePrePost{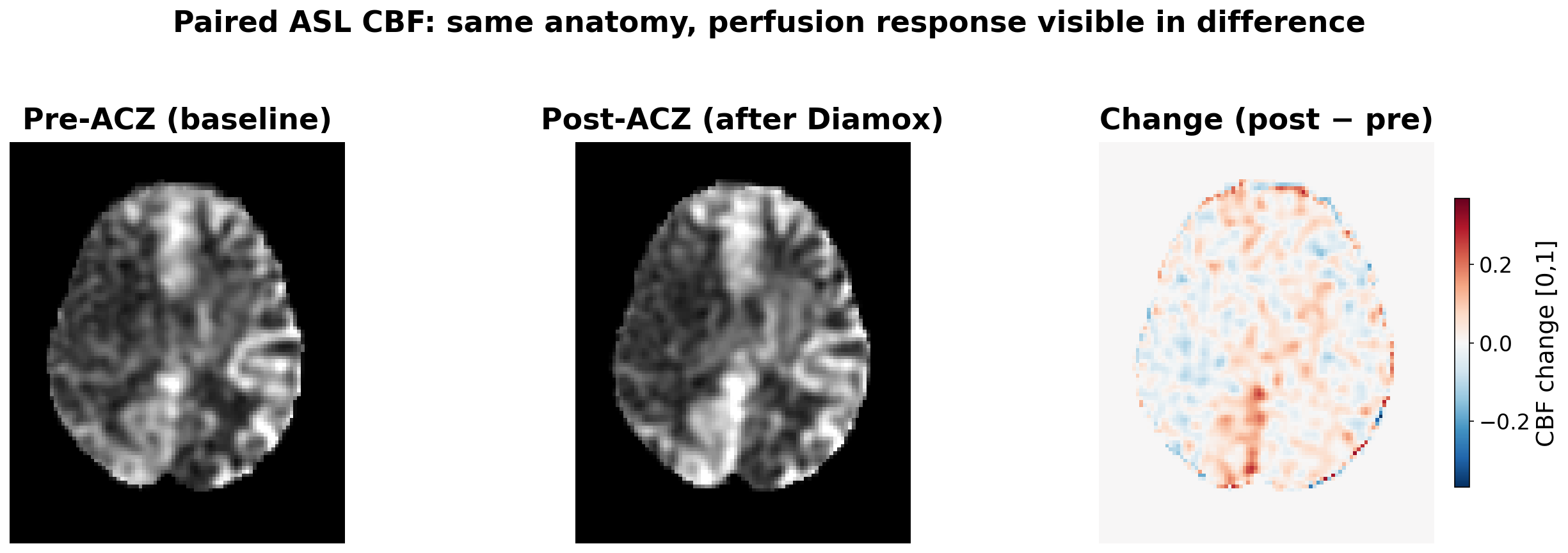}
 \gdef\FSlideFive{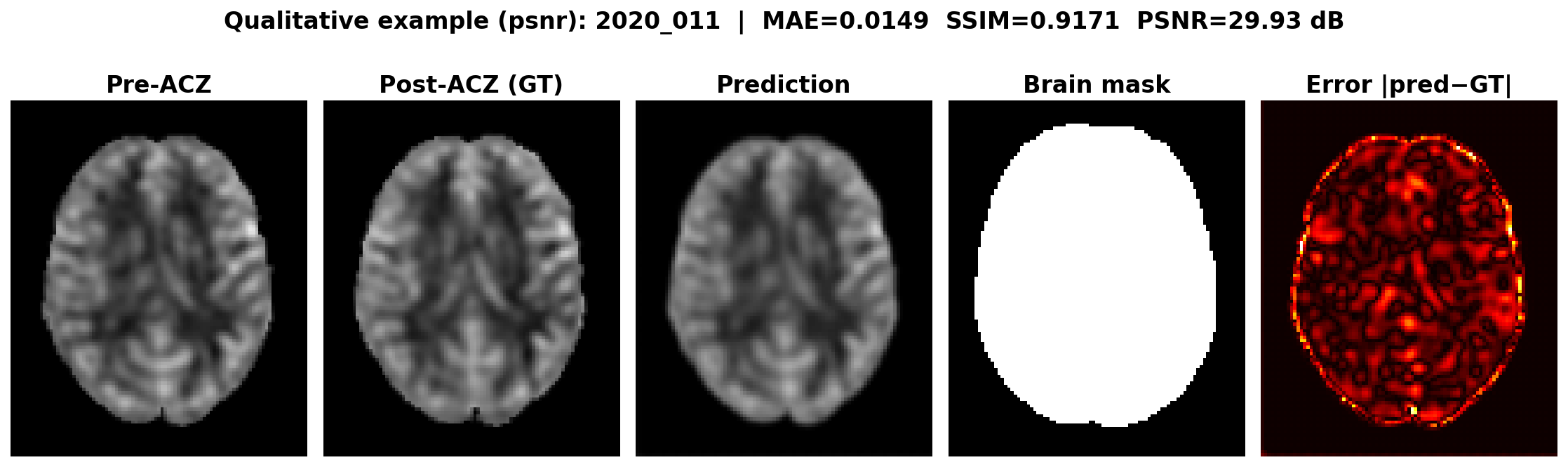}
 \gdef\FMaskCbf{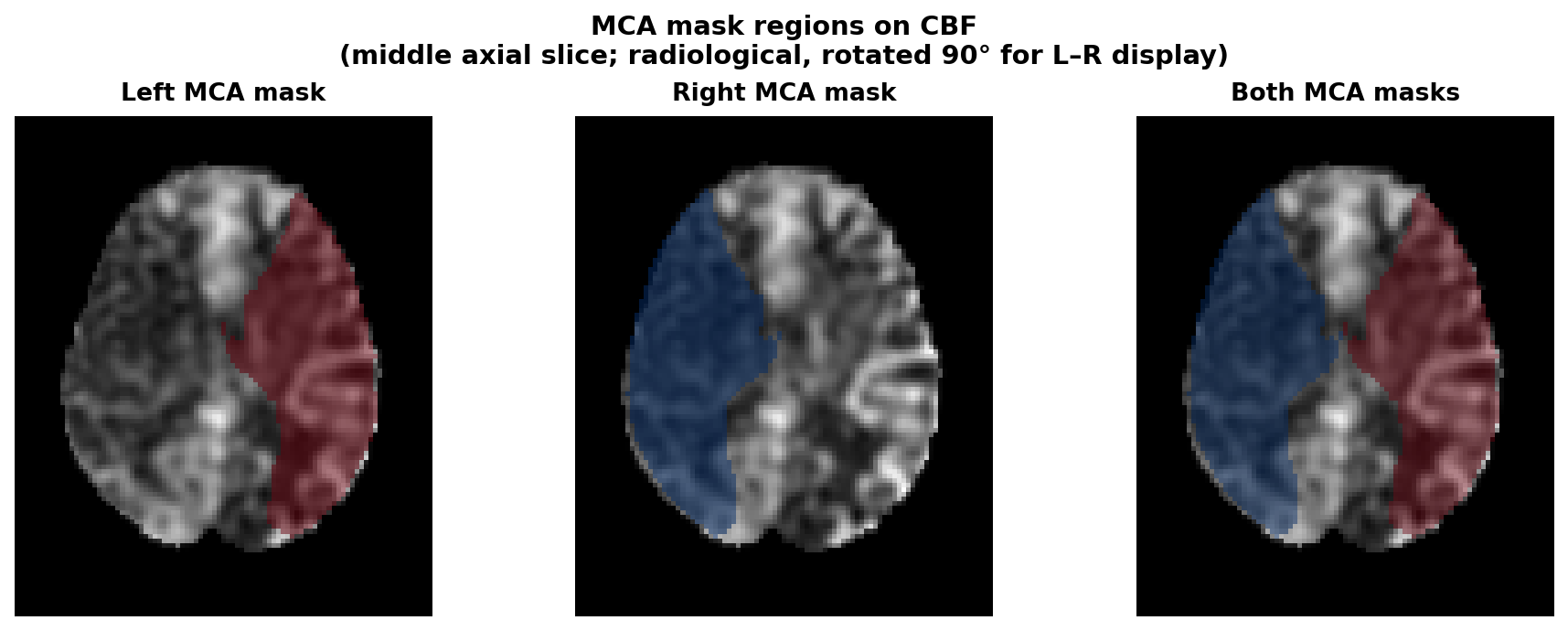}
 \gdef\FProofFig{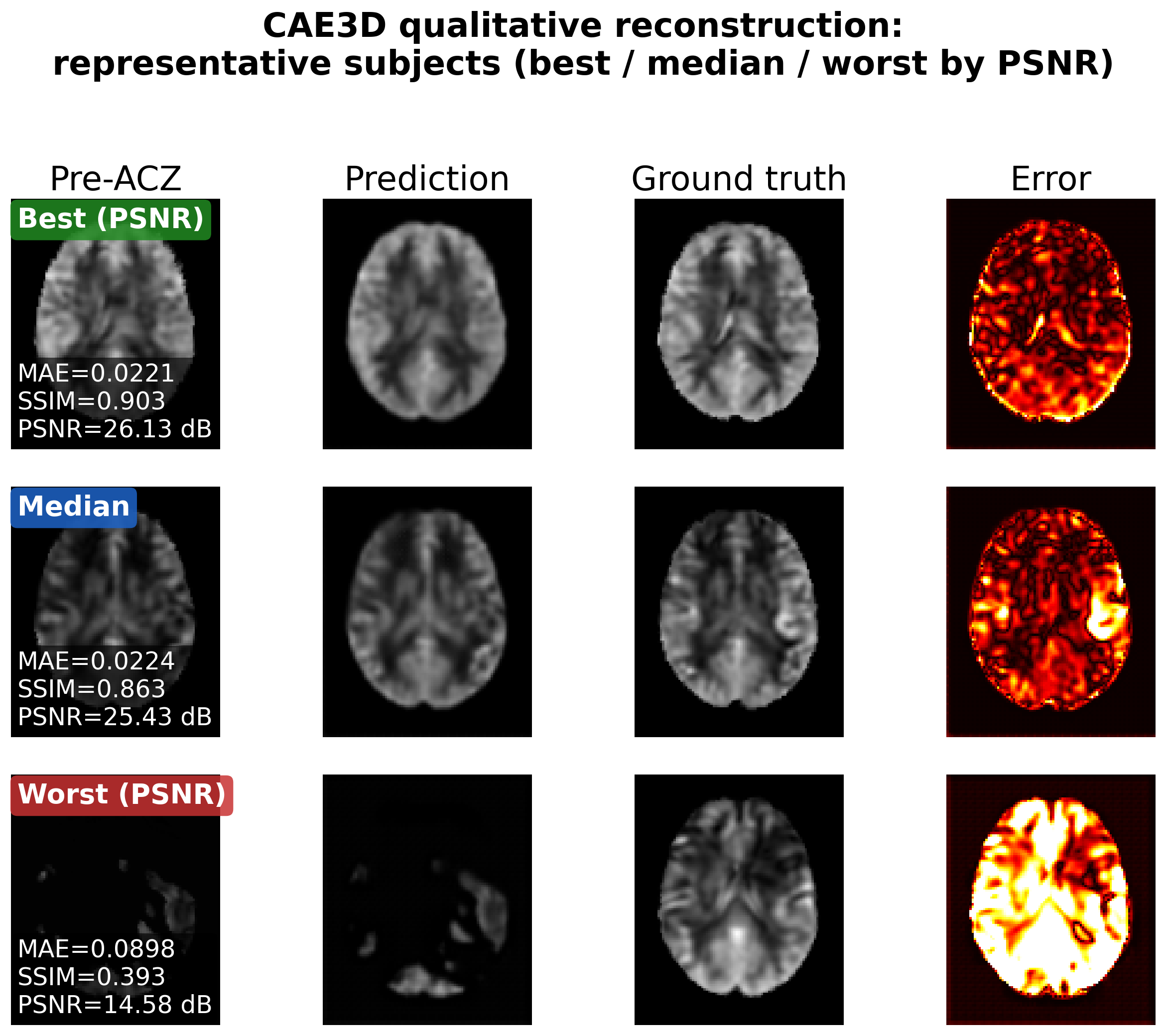}
 \gdef\FGuidePng{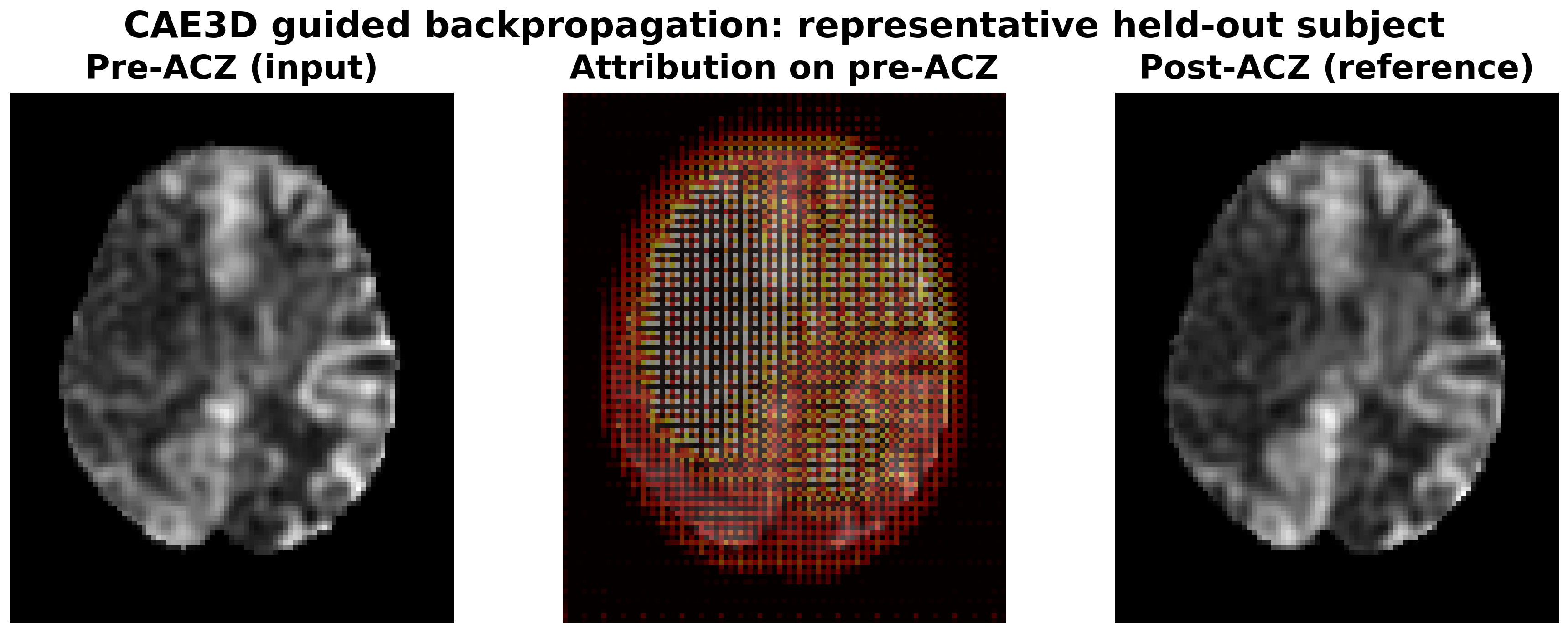}
 \gdef\FGuidePdf{guided_backprop_cae3d_panel.pdf}
 \gdef\FigDir{figures/}
 \endgroup
 \usepackage{adjustbox}
 \makeatletter
 \newcommand{\includegraphicsSlideVis}[2][]{\def\@svp{}%
   \edef\@tryLocal{#2}%
   \edef\@tryFig{\FigDir#2}%
   \expandafter\IfFileExists\expandafter{\@tryFig}{%
     \let\@svp\@tryFig
   }{%
     \expandafter\IfFileExists\expandafter{\@tryLocal}{%
       \let\@svp\@tryLocal
     }{}}%
   \ifx\@svp\@empty
     \fbox{\parbox{0.92\linewidth}{\centering\small Figure asset unavailable.}}%
   \else
     \def\@visopts{#1}\ifx\@visopts\@empty
       \includegraphics{\@svp}%
     \else
       \includegraphics[#1]{\@svp}%
     \fi
   \fi}
 \newcommand{\includegraphicsMasksOnCbf}[1][]{\def\@mcof{}%
   \edef\@mcoTryFig{\FigDir\FMaskCbf}%
   \expandafter\IfFileExists\expandafter{\@mcoTryFig}{%
     \let\@mcof\@mcoTryFig
   }{%
     \expandafter\IfFileExists\expandafter{\FMaskCbf}{%
       \let\@mcof\FMaskCbf
     }{}}%
   \ifx\@mcof\@empty
     \fbox{\parbox{0.92\linewidth}{\centering\small Figure asset unavailable.}}%
   \else
     \def\@visopts{#1}\ifx\@visopts\@empty
       \includegraphics{\@mcof}%
     \else
       \includegraphics[#1]{\@mcof}%
     \fi
   \fi}
 \newcommand{\ResolveSlideVis}[2]{%
   \def\@rsvresult{}%
   \edef\@rsvLocal{#1}%
   \edef\@rsvFig{\FigDir#1}%
   \expandafter\IfFileExists\expandafter{\@rsvFig}{%
     \let\@rsvresult\@rsvFig
   }{%
     \expandafter\IfFileExists\expandafter{\@rsvLocal}{%
       \let\@rsvresult\@rsvLocal
     }{}}%
   \global\let#2\@rsvresult
 }
 \makeatother
  \usepackage{booktabs}
 \usepackage{tikz}
 \usetikzlibrary{positioning,arrows.meta,calc}
 \usepackage{longtable}
\usepackage{float}
\usepackage{enumitem}
 \newcommand{\AtlasSliceFigWidth}{0.45\linewidth}
 \usepackage{siunitx}
 \usepackage[normalem]{ulem}

 \theorembodyfont{\upshape}
 \theoremheaderfont{\scshape}
 \theorempostheader{:}
 \theoremsep{\newline}

 \jmlrproceedings{PMLR}{Proceedings of Machine Learning Research}
 \jmlrvolume{340}
 \jmlryear{2026}
 \jmlrworkshop{Machine Learning for Healthcare}

\title[Predict Post-ACZ CBF in High-Risk Stroke Patients]{ Synthesizing Post-Acetazolamide Cerebral Blood Flow Maps from Baseline MRI in Moyamoya Using 3D Generative AI
}

\author{%
{\normalfont
\begin{tabular*}{\linewidth}{@{\extracolsep{\fill}}l r@{}}
\textbf{Julia Huang}\textsuperscript{1} & \textsc{julih@stanford.edu} \\
\textbf{Camila Gonzalez}\textsuperscript{2,3} & \footnotesize\textsc{camgonza@stanford.edu, camila.gonzalez@meduniwien.ac.at}\normalsize \\
\textbf{Rydham Goyal}\textsuperscript{1} & \textsc{rydham@stanford.edu} \\
\textbf{Aja Zou}\textsuperscript{4} & \textsc{ajazou@stanford.edu} \\
\textbf{Sasha Alexander}\textsuperscript{4} & \textsc{sashalex@stanford.edu} \\
\textbf{Michael Moseley}\textsuperscript{2} & \textsc{moseley@stanford.edu} \\
\textbf{Moss Y. Zhao}\textsuperscript{4}\thanks{These authors share senior authorship.} & \textsc{mosszhao@stanford.edu} \\
\textbf{Gary K. Steinberg}\textsuperscript{4,*} & \textsc{cerebral@stanford.edu} \\
\end{tabular*}\\[0.3em]
{\small\itshape
\textsuperscript{1}\,Department of Computer Science, Stanford School of Engineering, Stanford, CA 94305\\
\textsuperscript{2}\,Department of Radiology, Stanford School of Medicine, Stanford, CA 94305\\
\textsuperscript{3}\,Department of Anesthesia, Intensive Care Medicine, and Pain Medicine, Medical University of Vienna, Vienna, Austria 1090\\
\textsuperscript{4}\,Department of Neurosurgery, Stanford School of Medicine, Stanford, CA 94305
}
}
}
\jmlrauthors{J. Huang, C. Gonzalez, R. Goyal, A. Zou, S. Alexander, M. Moseley, M.Y. Zhao, and G.K. Steinberg}

\date{} 

\begin{document}
 \maketitle
\hypersetup{hypertexnames=false}

\begin{abstract}
For patients with Moyamoya disease, impaired cerebrovascular reserve (CVR) is an important hemodynamic criterion for recommending extracranial-to-intracranial bypass surgery, making reliable CVR assessment central to treatment planning. The reference protocol used in this cohort requires paired arterial spin labeling (ASL) perfusion MRI acquired before and after administration of the vasodilator acetazolamide (ACZ). ACZ may be contraindicated or avoided in patients with substantial renal dysfunction, relevant hypersensitivity, marked electrolyte derangement, or pregnancy, depending on clinical circumstances and institutional protocol. For affected patients in whom ACZ is not administered, the standard two-scan protocol cannot be completed as intended, and the post-ACZ cerebral blood flow (CBF) data used for hemodynamic assessment and bypass planning are unavailable. We propose \textbf{CAE3D}, a deterministic 3D conditional autoencoder that synthesizes post-ACZ CBF maps directly from pre-ACZ ASL input. Among eleven evaluated models (CAE3D, seven directly comparable full-volume in-house baselines spanning deterministic and diffusion-style variants, one middle-slice 2D contextual baseline (CAE\_2D), and two frozen-encoder foundation adapters reported separately as contextual references), CAE3D achieves the lowest held-out MAE (MAE~0.066, SSIM~0.80, PSNR~24.0\,dB) with near-zero full-brain mean bias, though its regional $\Delta$CBF predictions compress the dynamic range in high-response territories. Its MAE advantage was statistically significant (paired Wilcoxon, Holm-adjusted) over seven of the eight other trained-from-scratch baselines, with the exception of the 2D middle-slice CAE\_2D comparator; its SSIM and PSNR advantages were significant over all eight. These results establish the retrospective feasibility of post-ACZ CBF synthesis in patients who completed the standard two-scan protocol; extension to ACZ-contraindicated patients, who were not represented in this cohort, awaits external and prospective validation.

\end{abstract}
 \section{Introduction}
 \label{sec:intro}
 
Moyamoya disease is a rare, progressive cerebrovascular disorder in which the major intracranial arteries gradually narrow and occlude, raising the long-term risk of ischemic and hemorrhagic stroke~\citep{moyamoya}. Cerebrovascular reserve (CVR) reflects the brain's capacity to augment cerebral blood flow (CBF) in response to a vasodilatory stimulus. It is a key hemodynamic marker for disease severity in Moyamoya~\citep{AHAStrokeCVRMoyamoya2022}, and impaired CVR is an important hemodynamic criterion for recommending extracranial-to-intracranial bypass surgery over conservative management~\citep{AHAStrokeCVRMoyamoya2022,pmc9274857}.

In clinical practice, CVR is assessed using an acetazolamide (ACZ) challenge: paired pre- and post-ACZ ASL perfusion maps are an established clinical approach to CVR assessment~\citep{pmc9274857,AHAStrokeCVRMoyamoya2022}. However, ACZ may be contraindicated or avoided in patients with substantial renal dysfunction, clinically relevant hypersensitivity (including possible cross-sensitivity with sulfonamides), marked electrolyte imbalance, or pregnancy, depending on clinical circumstances and institutional protocol~\citep{acetazolamide, AHAStrokeCVRMoyamoya2022, pmc9274857}; this study's specific exclusion thresholds are given in \S\ref{sec:population}. These contraindication categories may arise in patients undergoing Moyamoya evaluation, although their prevalence cannot be estimated from the present cohort. Moreover, even when no formal contraindication exists, the two-scan protocol may be difficult to complete due to acute illness, scheduling, high cost, or visit-length constraints. In either case, the treating team is left without one of the key hemodynamic inputs used in bypass planning.

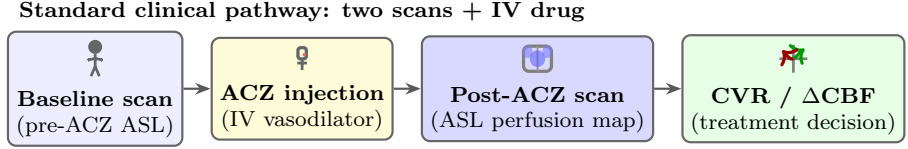
\begin{figure}[!t]
  \centering
  \begin{minipage}{0.78\linewidth}
    \centering
    \resizebox{\linewidth}{!}{%
       \begin{tikzpicture}[
         node distance=0.18cm and 0.35cm,
         iconbox/.style={rectangle, draw=black!60, thick, rounded corners=3pt,
                         minimum width=1.55cm, minimum height=0.85cm,
                         align=center, font=\scriptsize, inner sep=3pt},
         sublbl/.style={font=\scriptsize\itshape, align=center, text=black!70},
         hdrlbl/.style={font=\scriptsize\bfseries, align=center},
         arr/.style={-{Stealth[length=2mm]}, thick, black!65}
       ]
       \node[iconbox, fill=blue!7] (pat1) {%
         \tikz[scale=0.18]{\fill[black!60] (0,1.2) circle(0.5);
           \draw[black!60,line width=1pt](0,0.7)--(0,-0.2)
             (-0.4,0.3)--(0,0.5)--(0.4,0.3)
             (0,-0.2)--(-0.3,-0.9) (0,-0.2)--(0.3,-0.9);}\\[1pt]
         \textbf{Baseline scan}\\(pre-ACZ ASL)};
       \node[iconbox, fill=yellow!18, right=of pat1] (drug) {%
         \tikz[scale=0.18]{\draw[black!60,line width=1.2pt,rounded corners=1pt]
           (-0.3,0.8) rectangle (0.3,1.6);
           \draw[black!60,line width=1pt](0,0.8)--(0,0.0)
             (-0.35,0.5)--(0.35,0.5);
           \fill[red!60](0.15,1.1) circle(0.12);}\\[1pt]
         \textbf{ACZ injection}\\(IV vasodilator)};
       \node[iconbox, fill=blue!15, right=of drug] (scan) {%
         \tikz[scale=0.18]{\draw[black!50,line width=1.2pt,rounded corners=2pt]
           (-1,0.9) rectangle (1,-0.9);
           \draw[black!40,line width=0.7pt](0,0.9)--(0,-0.9);
           \fill[blue!50,opacity=0.6] (-0.7,0.5) ellipse (0.35 and 0.45);
           \fill[blue!50,opacity=0.6] (0.7,0.5) ellipse (0.35 and 0.45);
           \fill[blue!70,opacity=0.5] (0,0.0) ellipse (0.55 and 0.6);}\\[1pt]
         \textbf{Post-ACZ scan}\\(ASL perfusion map)};
       \node[iconbox, fill=green!10, right=of scan] (cvr) {%
         \tikz[scale=0.18]{\draw[black!50,line width=1pt](0,-1.0)--(0,1.0)
           (-1.0,0)--(1.0,0);
           \draw[green!60!black,line width=1.2pt,->](0,0)
             ..controls(0.3,0.2) and (0.5,0.5)..(0.8,0.7);
           \draw[red!60!black,line width=1.2pt,->](0,0)
             ..controls(-0.3,-0.1) and (-0.5,0.3)..(-0.8,0.6);}\\[1pt]
         \textbf{CVR / $\Delta$CBF}\\(treatment decision)};
       \draw[arr] (pat1) -- (drug);
       \draw[arr] (drug) -- (scan);
       \draw[arr] (scan) -- (cvr);
       \node[hdrlbl, above=0.06cm of drug]
         {\textbf{Standard clinical pathway}: two scans + IV drug};
       \end{tikzpicture}%
    }%
  \end{minipage}
  \caption{\textbf{Established clinical pathway.} The protocol used in this cohort requires two ASL scans (pre- and post-ACZ) bracketing an intravenous acetazolamide administration.}
  \label{fig:overview_standard}
 \end{figure}

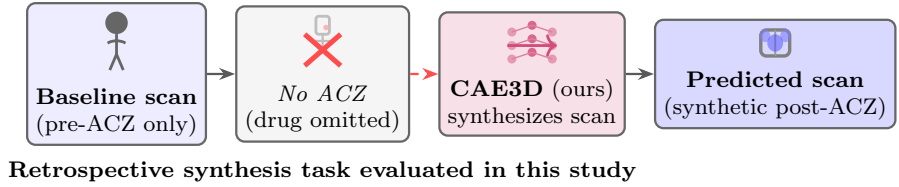
\begin{figure}[tb]
  \centering
  \begin{minipage}{0.78\linewidth}
    \centering
    \resizebox{\linewidth}{!}{%
       \begin{tikzpicture}[
         node distance=0.18cm and 0.35cm,
         iconbox/.style={rectangle, draw=black!60, thick, rounded corners=3pt,
                         minimum width=1.55cm, minimum height=0.85cm,
                         align=center, font=\scriptsize, inner sep=3pt},
         sublbl/.style={font=\scriptsize\itshape, align=center, text=black!70},
         hdrlbl/.style={font=\scriptsize\bfseries, align=center},
         arr/.style={-{Stealth[length=2mm]}, thick, black!65},
         xarr/.style={-{Stealth[length=2mm]}, thick, red!70, dashed}
       ]
       \node[iconbox, fill=blue!7] (pat2) {%
         \tikz[scale=0.28]{\fill[black!60] (0,1.2) circle(0.5);
           \draw[black!60,line width=1pt](0,0.7)--(0,-0.2)
             (-0.4,0.3)--(0,0.5)--(0.4,0.3)
             (0,-0.2)--(-0.3,-0.9) (0,-0.2)--(0.3,-0.9);}\\[2pt]
         \textbf{Baseline scan}\\(pre-ACZ only)};
       \node[iconbox, fill=gray!8, right=of pat2] (noacz) {%
         \tikz[scale=0.28]{
           \draw[black!25,line width=1pt,rounded corners=1pt]
             (-0.3,0.8) rectangle (0.3,1.6);
           \draw[black!25,line width=0.8pt](0,0.8)--(0,0.0)
             (-0.35,0.5)--(0.35,0.5);
           \fill[red!30](0.15,1.1) circle(0.12);
           \draw[red!70,line width=1.8pt](-0.8,-0.6)--(0.8,0.8);
           \draw[red!70,line width=1.8pt](0.8,-0.6)--(-0.8,0.8);}\\[2pt]
         \textit{No ACZ}\\(drug omitted)};
       \node[iconbox, fill=purple!12, right=of noacz, minimum width=1.6cm] (cae) {%
         \tikz[scale=0.28]{
           \foreach \y in {0.8,0.2,-0.4}{
             \fill[purple!40] (-0.9,\y) circle(0.2);
             \fill[purple!60] (0,\y+0.3) circle(0.2);
             \fill[purple!40] (0.9,\y) circle(0.2);
             \draw[purple!50,line width=0.5pt](-0.9,\y)--(0,\y+0.3);
             \draw[purple!50,line width=0.5pt](0.9,\y)--(0,\y+0.3);}
           \draw[purple!70,line width=1pt,->](-1.1,0.2)--(1.1,0.2);}\\[2pt]
         \textbf{CAE3D} (ours)\\synthesizes scan};
       \node[iconbox, fill=blue!15, right=of cae] (synth) {%
         \tikz[scale=0.18]{\draw[black!50,line width=1.2pt,rounded corners=2pt]
           (-1,0.9) rectangle (1,-0.9);
           \draw[black!40,line width=0.7pt](0,0.9)--(0,-0.9);
           \fill[blue!50,opacity=0.6] (-0.7,0.5) ellipse (0.35 and 0.45);
           \fill[blue!50,opacity=0.6] (0.7,0.5) ellipse (0.35 and 0.45);
           \fill[blue!70,opacity=0.5] (0,0.0) ellipse (0.55 and 0.6);}\\[1pt]
         \textbf{Predicted scan}\\(synthetic post-ACZ)};
       \draw[arr] (pat2) -- (noacz);
       \draw[xarr] (noacz) -- (cae);
       \draw[arr] (cae) -- (synth);
       \node[hdrlbl, below=0.10cm of noacz]
         {\textbf{Retrospective synthesis task evaluated in this study}};
       \end{tikzpicture}%
    }%
  \end{minipage}
  \caption{\textbf{Retrospective synthesis task evaluated in this study.} CAE3D predicts an independently normalized post-ACZ map from the pre-ACZ input. This diagram illustrates the potential future workflow motivation, not a clinically validated replacement for ACZ imaging.}
  \label{fig:overview_substitution}
 \end{figure}

To address this clinical gap, synthesizing post-ACZ CBF maps from baseline pre-ACZ ASL input using machine learning could offer a path toward CVR-relevant information without drug administration, pending validation. Encoder-decoder networks, diffusion models, and pretrained 3D encoders have each shown promise for this class of synthesis task in neuroimaging~\citep{kazerouni2023diffusionmodelsmedicalimage,OzbeySynDiff2023,GuoPredicting15OWaterPET2020,HusseinMedImageAnal2024}. \citet{goyal2026generating} established a 2D conditional synthesis approach for post-ACZ maps from pre-ACZ ASL. We extend that approach to full-volume 3D. We propose \textbf{CAE3D}, a 3D conditional autoencoder illustrated in Figures~\ref{fig:overview_standard}--\ref{fig:overview_substitution}, and evaluate it against ten comparators under the model-count convention defined in \S\ref{sec:methods} (eleven models evaluated in total, including CAE3D). The contribution of CAE3D lies in its task-specific adaptation and rigorous comparative evaluation of what is, to our knowledge, the first reported full-volume 3D study of this specific pre-ACZ-to-post-ACZ ASL synthesis task, rather than in a novel backbone architecture.

\paragraph{Contributions.}
This work makes two primary contributions. First, we demonstrate the feasibility of synthesizing post-ACZ CBF maps from pre-ACZ ASL input in patients with Moyamoya disease using 3D conditional image-synthesis models: \textbf{CAE3D} achieves the lowest held-out MAE among the nine in-house models in Table~\ref{tab:main_metrics} (\S\ref{sec:methods}), with near-zero Bland-Altman bias and lower territory-level $\Delta$CBF error than the evaluated diffusion baselines. Second, we extend the slice-based approach of \citet{goyal2026generating} to full-volume 3D; CAE3D outperformed the middle-slice \textbf{CAE\_2D} baseline on the reported metrics, though the two operate on different spatial domains and this comparison is supportive rather than a controlled dimensionality ablation. Implementation and evaluation scripts are available at \url{https://github.com/TheClassicTechno/cae3d-moyamoya-cbf-synthesis}.

\paragraph{Generalizable Insights about Machine Learning in the Context of Healthcare.}
For clinicians and ML researchers beyond this cohort, the results suggest that baseline-to-challenge image synthesis may warrant investigation in other clinical workflows involving paired baseline and pharmacologic-challenge acquisitions. Such extensions would require task-specific validation, particularly using agreement, regional, and subgroup-level metrics rather than global image-similarity scores alone. This single-center evaluation also offers three further methodological observations that may be relevant to related perfusion-synthesis settings, though not claimed to generalize beyond the evaluated models and protocols. First, deterministic 3D autoencoders performed better than the evaluated diffusion implementations in this dataset and training regime. Second, multimodal evaluation (voxel-level metrics with Bland-Altman agreement, bootstrap CIs, and vascular-territory $\Delta$CBF summaries) exposed limitations not visible in global metrics alone and is offered as a candidate template for other paired perfusion-synthesis tasks. Third, the tested frozen-encoder adaptations did not outperform task-specific training.

 \begin{figure}[tb]
  \centering
  \begin{minipage}{0.45\linewidth}
    \centering
    \resizebox{\linewidth}{!}{%
    \begin{tikzpicture}[
      node distance=0.28cm and 1.2cm,
      blk/.style={rectangle, draw=black!55, rounded corners=3pt,
                  minimum width=1.3cm, minimum height=0.40cm,
                  font=\scriptsize\sffamily, align=center, inner sep=2pt},
      enc/.style={blk, fill=blue!14},
      dec/.style={blk, fill=orange!14},
      bot/.style={blk, fill=purple!20, minimum width=1.2cm},
      ioblk/.style={blk, fill=gray!10},
      arr/.style={-{Stealth[length=2mm]}, thick, black!60},
      skip/.style={-{Stealth[length=1.5mm]}, dashed, blue!50, line width=0.7pt},
    ]
    \node[ioblk] (inp) {pre-ACZ\\(input)};
    \node[enc, below=0.28cm of inp] (e1) {Enc-1 / 16\,ch};
    \node[enc, below=0.28cm of e1]  (e2) {Enc-2 / 32\,ch};
    \node[enc, below=0.28cm of e2]  (e3) {Enc-3 / 64\,ch};
    \node[bot, below=0.28cm of e3]  (bot) {Bottleneck / 128\,ch};
    \node[ioblk, right=1.2cm of inp] (out) {pred.\ post-ACZ\\(output)};
    \node[dec, right=1.2cm of e1]   (d1) {Dec-1 / 16\,ch};
    \node[dec, right=1.2cm of e2]   (d2) {Dec-2 / 32\,ch};
    \node[dec, right=1.2cm of e3]   (d3) {Dec-3 / 64\,ch};
    \draw[arr] (inp) -- (e1);
    \draw[arr] (e1)  -- (e2);
    \draw[arr] (e2)  -- (e3);
    \draw[arr] (e3)  -- (bot);
    \draw[arr] (bot.east) -- ++(0.6cm,0) |- (d3.west);
    \draw[arr] (d3) -- (d2);
    \draw[arr] (d2) -- (d1);
    \draw[arr] (d1) -- (out);
    \draw[skip] (e3.east) -- node[above,font=\tiny,text=blue!70]{skip} (d3.west);
    \draw[skip] (e2.east) -- node[above,font=\tiny,text=blue!70]{skip} (d2.west);
    \draw[skip] (e1.east) -- node[above,font=\tiny,text=blue!70]{skip} (d1.west);
    \end{tikzpicture}%
    }%
  \end{minipage}
  \caption{\textbf{CAE3D architecture.} Skip-connected 3D encoder-decoder (three downsampling levels).}
  \label{fig:architecture}
 \end{figure}
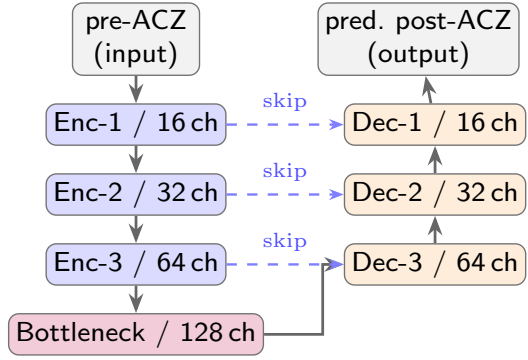

 \begin{figure}[tb]
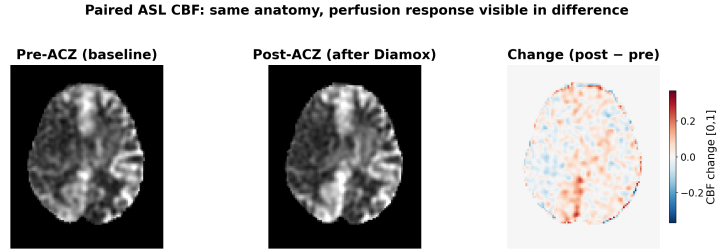

   \centering
   \begin{minipage}{0.62\linewidth}
     \centering
     \includegraphicsSlideVis[width=\linewidth]{\FSlidePrePost}
   \end{minipage}
   \caption{\textbf{Image context.} Pre-ACZ, post-ACZ GT, and post-minus-pre change (middle axial slice, one subject). Image panels use intensity windowing for visualization ($v_{\max}$ set to the 99th percentile of non-zero voxels); see the plotting scripts in the supplementary material for exact details.}
   \label{fig:image_context}
 \end{figure}

 \begin{figure}[tb]
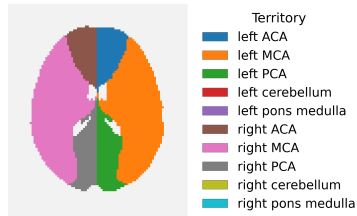

   \centering
   \begin{minipage}{\AtlasSliceFigWidth}
     \centering
     \includegraphicsSlideVis[width=\linewidth]{\FSlideVascular}
   \end{minipage}
   \caption{\textbf{Vascular territory atlas}~\citep{LiuVascularAtlasScientificData} on a representative pre-ACZ slice; defines regional summaries (\S\ref{sec:metrics}). Windowed for visualization as in Figure~\ref{fig:image_context}.}
   \label{fig:vascular_territories_colormap}%
 \end{figure}

 \section{Related Work}
 \label{sec:related}

\paragraph{Background: ASL perfusion imaging.}
Arterial spin labeling (ASL) is a non-invasive MRI technique that uses magnetically labeled arterial blood water as an endogenous tracer to quantify cerebral blood flow (CBF) without exogenous contrast. Each ASL acquisition yields a voxelwise CBF map. The change between pre- and post-ACZ CBF maps, $\Delta\text{CBF} = (\text{CBF}_\text{post} - \text{CBF}_\text{pre})/\text{CBF}_\text{pre} \times 100\,\%$, encodes the regional vasodilatory response used in CVR assessment and may contribute to bypass planning. The synthesis task in this work is therefore image-to-image regression: predicting the post-ACZ CBF map from the pre-ACZ CBF map in normalized voxel space, where each volume is independently rescaled to $[0,1]$ (\S\ref{sec:data}); CAE3D synthesizes post-ACZ maps in this independently normalized intensity space rather than in physical CBF units, which bears on how normalized-space $\Delta$CBF relates to physiological CVR (\S\ref{sec:discussion_limitations}).

\paragraph{CVR assessment with acetazolamide administration.}
Clinicians assess hemodynamic reserve in Moyamoya using pharmacologic vasodilation with CVR imaging~\citep{AHAStrokeCVRMoyamoya2022,pmc9274857}; the practical constraints (time, logistics, cost, contraindications) that motivate computational alternatives are detailed in \S\ref{sec:intro}.
 
\paragraph{Post-ACZ perfusion synthesis from baseline ASL perfusion maps.}
\citet{goyal2026generating} established slice-based conditional and diffusion baselines for this clinical problem, focusing on a 2D middle-slice prediction with a smaller model comparison. By contrast, the present study evaluates full-volume 3D prediction, includes side-by-side 2D and 3D baselines within a single protocol, and extends the evaluation with agreement and territory-level analyses. The objective (synthetic post-ACZ from pre-ACZ alone) is the same.

\paragraph{Physiological basis and scope of the learned mapping.}
CVR is, by definition, the response to a vasodilatory challenge, so it is fair to ask why a pre-ACZ baseline map should carry any information about it. In Moyamoya, chronic large-vessel occlusion drives collateral recruitment and territorial steal well before ACZ is administered, so resting-state (pre-ACZ) perfusion already reflects, in part, how exhausted or preserved a territory's autoregulatory reserve is~\citep{AHAStrokeCVRMoyamoya2022,pmc9274857}. This motivates a statistical association between baseline perfusion patterns and vasodilatory response at the cohort level. We do not claim that two patients with visually identical pre-ACZ maps necessarily share identical CVR, nor that the model recovers an individual patient's exhausted-but-compensated reserve from the baseline scan alone; such cases are, by construction, the hardest for any baseline-only predictor. Rather, CAE3D learns the population-level correlation between baseline CBF patterns and average post-ACZ response within this Moyamoya cohort, not subject-specific CVR recovery. We return to the evidence for this distinction, including the modest per-subject $R^2$ and territory-level $\Delta$CBF compression, in \S\ref{sec:discussion}.

\paragraph{Broader perfusion and multimodal synthesis.}
Learning-based synthesis of perfusion-related targets from baseline imaging has also been explored in MRI--PET settings~\citep{HusseinMedImageAnal2024,HusseinarXivMRI2PET,DayarathnaSurveyMRICTPET2024}. Foundation models pretrained on large medical imaging datasets are increasingly applied to synthesis tasks; we include two publicly released 3D encoders as frozen-encoder comparators, \textbf{Med3DVLM}~\citep{Med3DVLM2025} and \textbf{SAM-Med3D}~\citep{SAMMed3D2023}, with adaptation details in \S\ref{sec:foundation_baselines}.

 \paragraph{3D modeling, evaluation rigor, and clinical interpretability.}
  Diffusion and cold-diffusion models have shown strong results in medical image synthesis~\citep{HoDDPM2020,BansalColdDiffusion2023}. Here, we provide a unified 3D comparison within a single evaluation framework (\S\ref{sec:methods}), including atlas-based regional summaries in addition to global image-quality metrics. Our in-house models build on standard MONAI~\citep{cardoso2022monai} encoder-decoder and diffusion components; implementation details are described in \S\ref{sec:proposed}.

 \section{Methods}
 \label{sec:methods}

\paragraph{Model-count convention.} We evaluated 11 models: CAE3D, eight in-house baselines, and two separately evaluated foundation-encoder adapters (Med3DVLM and SAM-Med3D). Of the eight in-house baselines, seven are full-volume 3D models directly comparable to CAE3D; the eighth, \textbf{CAE\_2D}, is evaluated on the middle slice only and is a descriptive, different-domain comparator (\S\ref{sec:baselines}). Primary held-out comparisons (Table~\ref{tab:main_metrics}) involve all nine in-house models (CAE3D plus the eight baselines), while CAE3D has ten total comparator rows (the eight in-house baselines plus the two foundation adapters). The three-seed stability table (Table~\ref{tab:fixed_test_three_seed}) and the $K$-fold table (Table~\ref{tab:kfold_rotating}, Appendix~C) report different, smaller model subsets tailored to those specific analyses, each listed explicitly in its own caption; \emph{nine}, \emph{ten}, and \emph{eleven} refer to the primary held-out comparison unless stated otherwise.

 \subsection{Proposed method: 3D conditional autoencoder and diffusion variants}
 \label{sec:proposed}
\textbf{CAE3D} is a 3D conditional autoencoder that maps a pre-ACZ perfusion map to a predicted post-ACZ perfusion map in a single forward pass.
CAE3D treats the task as deterministic conditional regression, prioritizing stable full-volume reconstruction rather than generation of multiple possible outputs.
Given the modest cohort size, a deterministic encoder-decoder also favors training stability and spatial fidelity.
CAE3D uses MONAI's volumetric encoder-decoder with skip connections between encoder and decoder stages, configured with three spatial dimensions, channel widths of 16, 32, 64, and 128, strides of 2 at each of the three downsampling levels, and 2 residual units per block, preserving fine-grained spatial detail while learning the pre$\rightarrow$post perfusion mapping (Algorithm~\ref{alg:cae3d}, steps 1-2). The output is a continuous perfusion intensity map; the model is termed a conditional autoencoder to distinguish it from segmentation networks. The training objective is
 \begin{equation}
 L(\theta) = L_{\mathrm{L1}}(\hat{x}, x) + \bigl(1 - \mathrm{SSIM}(\hat{x}, x)\bigr),
 \label{eq:cae3d}
 \end{equation}
where $\hat{x}$ is the predicted post-ACZ map, $x$ is the ground-truth post-ACZ map, $L_{\mathrm{L1}}$ is the mean absolute error, and SSIM is the structural similarity index (3D volumetric SSIM over the brain with data range 1.0; same settings for loss and evaluation). CAE3D was trained under the same optimizer and schedule as the other in-house models (\S\ref{sec:training}), without early stopping (full pseudocode: Algorithm~\ref{alg:cae3d}, Appendix~A). CAE3D loss ablation (L1-only / SSIM-only / L1+SSIM) is in the supplementary material; L1+SSIM matches the main row and yields the best MAE/PSNR. We also implemented an exploratory vascular-territory-weighted loss~\citep{LiuVascularAtlasScientificData}; all primary results reported here use the unweighted full-brain objective (in-mask voxels via the MNI brain mask; \S\ref{sec:metrics}).

\paragraph{Architecture details.}
\textbf{CAE3D} and several in-house encoder-decoder baselines are built on MONAI's 3D \texttt{UNet}~\citep{cardoso2022monai}, a standard implementation of the encoder-decoder U-Net design~\citep{RonnebergerUNet2015}. The contribution of CAE3D lies in the task-specific adaptation: the choice of spatial dimensions, channel widths, strides, residual units, inputs and targets, padding strategy, pre$\rightarrow$post mapping, loss design (the combined L1+SSIM objective, Eq.~\eqref{eq:cae3d}), and training protocol (full 50-epoch schedule selected by validation PSNR).

\paragraph{Protocol.} We studied supervised synthesis of post-ACZ ASL perfusion maps from pre-ACZ maps as the shared prediction task. We compared deterministic encoder-decoder models, diffusion-style variants, a Fourier Neural Operator baseline (FNO\_3D)~\citep{LiFNOICLR2021}, and adapted pretrained 3D encoders under this shared task. We used a fixed subject-level train/validation/test split (random seed 42, determined before any model was trained) for all models, with three training seeds (42, 123, 456) for in-house models and a held-out test set reserved for final evaluation (never used for tuning). \textbf{Secondary $K$-fold.} Separately from the primary held-out, we ran a rotating-test $K$-fold ($K{=}5$) over the full cohort for nine trained-from-scratch 3D models, including CAE3D and eight comparison models (\S\ref{sec:kfold_textbook}; replication: Appendix~C). Foundation adapters (Med3DVLM, SAM-Med3D) used the same combined split with seed-aggregated metrics (\S\ref{sec:foundation_baselines}).
 
\paragraph{Training.}
\label{sec:training}
All in-house models were trained with the Adam optimizer (learning rate $10^{-3}$), batch size 2, for up to 50 epochs on one NVIDIA GPU, with checkpoints selected by the highest validation PSNR. Most deterministic baselines converged by epoch~25--30; diffusion-style models (DDPM\_3D, Cold\_3D, Residual\_3D) ran the same epoch budget but required substantially more compute per epoch because of the full $T{=}1000$-step diffusion pass per training sample (early-stopping details for CAE3D vs.\ CAE3D-ES are in \S\ref{sec:baselines}). Loss terms were model-specific, but all models shared the same pre$\rightarrow$post prediction target and metric definitions; full-volume models were evaluated over the full brain volume, whereas CAE\_2D was evaluated on the middle axial slice. CAE-family models optimized L1+SSIM (\S\ref{sec:proposed}). Foundation-adapter training used the settings in \S\ref{sec:foundation_baselines}. Three randomized seeds (42, 123, 456) were run per in-house model, with test-set mean$\pm$std reported across seeds.
 
\paragraph{Evaluation.} The validation set was used only for model selection (checkpointing by validation PSNR) and early stopping where enabled; the held-out test set ($N=32$) was used only for final reporting. For in-house models, we reported mean$\pm$std of test metrics across $R{=}3$ independently trained seeds (Table~\ref{tab:fixed_test_three_seed}), which characterizes seed-to-seed training variability. Table~\ref{tab:main_metrics} instead reports, per in-house model, the bootstrap 95\% CI of the cohort mean from the predesignated seed-42 instance's per-subject metrics; within that seed, the checkpoint with the highest validation PSNR was retained, with no test-set involvement in selection ($N=32$ paired observations; the same per-subject values are used for the Wilcoxon tests; the selection rule is summarized in Table~\ref{tab:primary_checkpoints}, Appendix~A). Because Table~\ref{tab:main_metrics} and Table~\ref{tab:fixed_test_three_seed} are computed from different evaluation runs (a single instance vs.\ three independently trained seeds), their point estimates for a given model are not directly interchangeable. Bootstrap 95\% CIs for cohort-mean MAE, SSIM, and PSNR used $B{=}2000$ subject-level resamples (percentile method), quantifying sampling uncertainty of the cohort means rather than per-voxel variability. Bias CIs in Table~\ref{tab:main_metrics} used the same bootstrap on per-subject Bland-Altman differences (predicted minus target mean intensity); Limits of Agreement (LoA) are full-sample point estimates ($\pm 1.96\times$ Standard Deviation (SD)). We reported paired Wilcoxon tests with Holm adjustment as defined and detailed in \S\ref{sec:agreement} and the supplementary material.
 
\paragraph{Patient study population.}
\label{sec:population}
Moyamoya disease is exceptionally rare, with an estimated US incidence of 0.57 per 100,000 person-years~\citep{MillerNeurology2025}. The rarity of Moyamoya makes large paired cohorts difficult to assemble; nevertheless, the single-center design remains an important limitation (\S\ref{sec:discussion_limitations}). The study enrolled patients with Moyamoya disease who had not undergone prior neurosurgical treatment and were undergoing evaluation for extracranial-to-intracranial bypass surgery at the Stanford School of Medicine, which maintains one of the largest single-center Moyamoya cohorts in the Western Hemisphere. Inclusion required confirmation of Moyamoya disease by catheter cerebral angiography, MR angiography, or CT angiography, and completion of paired pre- and post-acetazolamide ASL perfusion maps. Exclusion criteria for the PET/MRI procedures included kidney function impairment (glomerular filtration rate $< 40$\,ml/min/1.73\,m\textsuperscript{2}), pregnancy, history of brain injury, and contraindications to MRI or to acetazolamide. At imaging, subjects had no acute infarction or hemorrhage. The study was approved by the local institutional review board (IRB), and all participants provided written consent. Data were acquired between 2020 and 2023 under an approved simultaneous PET/MRI ASL protocol; only the ASL perfusion (MRI) channel was used in this analysis, and PET data were not modeled. From the 252 subject-level pairs initially identified, one subject was excluded after failed affine registration (\S\ref{sec:data}). The final cohort consists of 251 subjects (188 for training, 31 for validation, and 32 for held-out test), all with Moyamoya disease.
 
\paragraph{Data and preprocessing.}
\label{sec:data}
All in-house models used the same subject split, registered and normalized input-target pairs, and metric definitions, preprocessed identically before any model-specific training (exceptions noted above, \S\ref{sec:training}--\S\ref{sec:foundation_baselines}).
Inputs and targets are ASL perfusion maps (pre-ACZ baseline, post-ACZ after ACZ administration), modeled as image-to-image regression in a normalized space. \textbf{Registration and normalization.} Volumes were affinely registered to MNI152 2\,mm (a standard whole-brain template space that enables cross-subject spatial comparison) with DIPY (center-of-mass, translation, rigid, affine; mutual information) and resampled to $91\times109\times91$~\citep{MNI152Collins1994,ICBM152Mazziotta2001}. The pre-ACZ transform was applied to both pre- and post-maps, and one failed case (251/252 success) was excluded. Trilinear interpolation was used for images and nearest-neighbor for masks; an MNI brain mask was applied, and per-volume min–max normalization to $[0,1]$ was performed independently for each pre- and post-ACZ volume (each volume rescaled to its own $[0,1]$ range). Because this rescaling is independent per volume, absolute between-scan intensity differences are partly absorbed into it, which bears on how normalized-space $\Delta$CBF relates to physical-unit CVR (\S\ref{sec:discussion_limitations}). No additional subject-level exclusions were applied beyond this single registration failure; the held-out test set of 32 subjects passed all preprocessing steps without further selection.

\paragraph{Model inputs and augmentation.} 2D baselines used the middle axial slice ($91\times109$), while 3D models used the full volume $(1,91,109,91)$. After MNI resampling, all volumes share this same $91\times109\times91$ grid. Some architectures require spatial dimensions divisible by $2^{\mathrm{depth}}$ (here, 8 for three downsampling levels); since 91, 109, and 91 are not divisible by 8, we zero-padded to $96\times112\times96$ for those architectures and cropped back to $91\times109\times91$ before computing all metrics. Training used paired random flips along the left-right (LR), anterior-posterior (AP), and superior-inferior (SI) axes along with paired intensity scaling in $[0.9, 1.1]$, with the same flip and scale applied to both volumes in a pair to preserve pre/post alignment. Only the LR flip is anatomically motivated by approximate brain symmetry; the AP and SI flips are not anatomically realistic and were included as generic geometric regularization rather than physiologically motivated augmentations. For the ResNet~3D model, we also evaluated Tied-Augment~\citep{KurtulusICML2023} (scheduled pooled-feature tie across two augmented views). This augmentation system improved validation metrics but did not surpass the best 3D conditional autoencoder on held-out testing.

  \subsection{Baselines and comparison models}
 \label{sec:baselines}
\paragraph{Setup and CAE naming.} The in-house models were compared using the shared subject split, preprocessing pipeline, and metric definitions, with the spatial-domain and foundation-adapter exceptions described below. The proposed encoder-decoder is \textbf{CAE3D (ours)} in tables. CAE3D and the following in-house baselines and ablations---CAE3D-ES, ResNet\_3D, Cold\_3D, Residual\_3D, DDPM\_3D, FNO\_3D, Hybrid\_3D, Patch\_3D---were implemented and trained under the shared protocol using three random seeds (besides the two foundation model adapters in \S\ref{sec:foundation_baselines}); CAE\_2D is introduced separately below. \textbf{CAE3D} uses the full training schedule \textbf{without} early stopping (50 epochs, checkpoint = highest validation PSNR); \textbf{CAE3D-ES} is the same encoder-decoder trained \textbf{with} early stopping when validation PSNR plateaus for 10 consecutive epochs. Both use the same MONAI implementation and L1+SSIM objective, differing only in the training-stop rule; CAE3D-ES is retained as an ablation to isolate that rule's effect. Its consistently higher MAE relative to CAE3D (Table~\ref{tab:fixed_test_three_seed}) supports the full 50-epoch schedule for this cohort size; no deterioration in the selected validation metric was observed.

\paragraph{Slicing, ResNet, Diffusion, and FNO.} \textbf{CAE\_2D} uses the same L1+SSIM loss on the middle axial slice only (conservative baseline; primary comparisons are 3D). \textbf{ResNet\_3D} is a 3D ResNet (MONAI) with the same loss and preprocessing; we also report a Tied-Augment variant~\citep{KurtulusICML2023}. \textbf{Diffusion models} concatenate pre-ACZ to the noisy or degraded target channel-wise; the denoiser matches the encoder-decoder resolution hierarchy (CAE3D family).\\ 
\texttt{DDPM\_3D}: $T{=}1000$, linear $\beta$, $\varepsilon$-prediction; full DDPM sampling (no DDIM). \\
\texttt{Cold\_3D}: linear interpolation from post-ACZ toward pre-ACZ ($x_t = \alpha_t x_{\text{post}} + (1-\alpha_t) x_{\text{pre}}$) with a cosine schedule; the network reverses this path. \\
\texttt{Residual\_3D}: diffuse and predict residual $\mathrm{post}-\mathrm{pre}$, then $\mathrm{pre}+\text{prediction}$; $T{=}1000$, linear $\beta$. \\
\texttt{Patch\_3D}: a patch-based latent diffusion variant; a VAE first encodes overlapping $24^3$ patches (stride 12, 50\% overlap) into a compact latent space, a diffusion model is trained to denoise in that latent space conditioned on the pre-ACZ patch, and patch predictions are decoded and re-assembled (overlap-averaged) into the full volume. \\

    Residual\_3D\_tips applies additional training-stabilization measures (DDIM sampling, patch-based training, a cosine noise schedule) intended to reduce the collapse behavior seen in \textbf{Residual\_3D}; on this cohort it did not succeed (\S\ref{sec:baselines}) and is not a primary row in Table~\ref{tab:main_metrics}. \textbf{FNO\_3D}~\citep{LiFNOICLR2021} is a Fourier Neural Operator from pre-ACZ to post-ACZ (12 modes, width 64) with the same L1+SSIM loss and pipeline. Diffusion and FNO models used the same held-out test set and full-brain metrics.

\subsection{Foundation model adapters: Med3DVLM and SAM-Med3D}
 \label{sec:foundation_baselines}
To contextualize trained-from-scratch results against transfer learning, two open-source 3D encoders were adapted to the same pre$\rightarrow$post ASL task, split, and full-brain evaluator: \textbf{Med3DVLM}~\citep{Med3DVLM2025} (frozen DCFormer encoder, 8 seeds) and \textbf{SAM-Med3D}~\citep{SAMMed3D2023} (frozen SAM 3D image encoder, 3 seeds), each paired with a small trainable regression decoder. ``Frozen'' here means that the pretrained encoder weights are held fixed; only the lightweight decoder head is trained on the ASL synthesis task. The regression decoder is a stack of \texttt{ConvTranspose3d} upsampling blocks (BatchNorm, GELU) that double spatial resolution until the target volume size is reached, followed by a final $1\times1\times1$ convolution to a single output channel; no skip connections are used, since the frozen backbones lack a compatible encoder-decoder feature hierarchy. Both are third-party, public architectures, distinct from the in-house MONAI baselines. Full training configurations and aggregated outputs are in Appendix~C.
 
 \subsection{Metrics}
 \label{sec:metrics}
MAE, SSIM, and PSNR use the same metric definitions for every model~\citep{ImageQualityPSNRSSIMIEEE}, over in-mask voxels (MNI brain mask) on $[0,1]$ data. For 3D models, metrics are computed over the full volume; \textbf{CAE\_2D} uses the middle axial slice only (\S\ref{sec:methods}). Foundation adapters (\textsuperscript{$\dagger$}) use the same evaluator but a separately described training/aggregation protocol (\S\ref{sec:foundation_baselines}). \textbf{MAE} refers to mean absolute error over in-mask voxels; lower MAE means closer voxel-wise intensity reconstruction. \textbf{SSIM} measures three-dimensional structural similarity (window size 7, data range 1.0), matching the loss; higher SSIM means better preservation of structural contrast and local spatial patterns. \textbf{PSNR} is the standard decibel (dB) measure derived from MSE; higher PSNR indicates lower reconstruction noise. We also report \textbf{$R^2$}, computed as $1 - \mathrm{SS}_{\mathrm{res}}/\mathrm{SS}_{\mathrm{tot}}$ on per-subject mean in-mask intensities (32 subjects) rather than on voxelwise values. We report all four metrics for eleven models in Table~\ref{tab:main_metrics}.

\paragraph{Regional perfusion and $\Delta$CBF.} We additionally evaluated performance across predefined vascular territories. Our regional pipeline covers nine 3D models, evaluated on the same 32-subject held-out set. Atlas masks from \citet{LiuVascularAtlasScientificData} at MNI 2\,mm define 11 regions (bilateral ACA, MCA, PCA, cerebellum, pons/medulla, plus a pooled vascular mask). Figure~\ref{fig:vascular_territories_colormap} shows the territory color map on a representative slice; Supplementary Figure~S1 shows mask unions (left hemisphere, right hem., pooled vascular) on the same slice convention. Per territory, all voxels within the atlas mask are first averaged to obtain territory-level scalars $\bar{I}_{\text{pre}}$, $\bar{I}_{\text{post}}^{\text{GT}}$, and $\bar{I}_{\text{post}}^{\text{pred}}$; percent change is then $\Delta\text{CBF} = (\bar{I}_{\text{post}} - \bar{I}_{\text{pre}})/\bar{I}_{\text{pre}} \times 100\,\mbox{\%}$, and absolute change is $\Delta\text{CBF}_{\text{abs}} = \bar{I}_{\text{post}} - \bar{I}_{\text{pre}}$ (normalized units). Computing on territory means rather than voxel-wise ratios avoids denominator instability caused by individual near-zero-perfusion voxels. Laterality metadata for stratification are unavailable; this analysis is not a formal gas-reactivity CVR (\S\ref{sec:discussion}).
 
\paragraph{Agreement and statistical significance.}
\label{sec:agreement}
Bland-Altman agreement analysis~\citep{BlandAltman1986} was used to assess whether predicted and target perfusion summaries agree closely enough for interpretation, by quantifying systematic bias and the spread of paired differences rather than relying solely on correlation. For each model, we computed the mean difference (predicted minus target) and LoA ($\pm 1.96\times$ SD) from per-subject full-brain mean intensities. Agreement should be interpreted together with LoA and reconstruction errors, rather than solely from bias. Table~\ref{tab:bland_altman} duplicates bias/LoA. Pairwise comparisons used the paired Wilcoxon signed-rank test on per-subject MAE, SSIM, and PSNR, with Holm adjustment to control for multiple comparisons within each metric; full tables are in the supplementary material. Figure~\ref{fig:guided_backprop} illustrates CAE3D guided backpropagation (input-gradient saliency maps highlighting which pre-ACZ voxels most influence the prediction; positive-gradient variant; full set in the supplementary material); representational similarity analysis (RSM: pairwise cosine similarities between bottleneck features across test subjects) is reported in the supplementary material only.

\section{Results}

\paragraph{Primary test-set reconstruction.}
CAE3D achieved the lowest MAE (0.066) among all trained-from-scratch models on the held-out test set, and tied for the highest reported SSIM (0.80, with ResNet\_3D) and PSNR (24.0\,dB, with FNO\_3D) at the precision shown in Table~\ref{tab:main_metrics}.
Published pCASL test-retest studies report within-subject coefficients of variation of roughly 3--8\% for regional CBF under a fixed acquisition protocol~\citep{LinASLTestRetest2020,NeumannASLTestRetest2021}. Because our MAE is computed voxelwise after independent per-volume $[0,1]$ normalization, it is not directly comparable in units or normalization to these physical-unit regional repeatability estimates; we report the test-retest figures only as qualitative context for the scale of ASL measurement variability, and we do not claim that MAE~0.066 is below, or otherwise directly comparable to, the acquisition's own repeat-scan noise floor.
Table~\ref{tab:main_metrics} reports cohort-mean MAE, SSIM, and PSNR with bootstrap confidence intervals, together with per-subject $R^2$ and Bland-Altman bias/limits of agreement.
For the adapted pretrained rows (\textsuperscript{$\dagger$}), MAE/SSIM/PSNR are mean$\pm$std over seeds, and $R^2$ and bias/LoA are point estimates from one checkpoint (\S\ref{sec:foundation_baselines}, Appendix~C).
Five-fold cross-validation (Table~\ref{tab:kfold_rotating}, Appendix~C; \S\ref{sec:kfold_textbook}) confirmed CAE3D remained top-performing across alternative partitions.

 \begin{table}[tb]
 \centering
\caption[Primary held-out test metrics]{Cohort mean MAE $\downarrow$, SSIM $\uparrow$, PSNR $\uparrow$; $R^2$ and Bland-Altman bias/LoA. Values shown as mean $\pm$ half-width of bootstrap 95\% CI ($B{=}2000$). \textbf{Bold}/underline: best/second-best in-house model (unrounded values; displayed ties are a rounding artifact, \S\ref{sec:results_holdout}). \textsuperscript{$\dagger$}~Foundation adapter rows (\S\ref{sec:foundation_baselines}) use a different protocol; bold/underline compares only Med3DVLM vs.\ SAM-Med3D ($\pm$ denotes std over seeds). See Appendix~C for foundation $R^2$/bias/LoA.}
 \label{tab:main_metrics}
 \footnotesize
 \setlength{\tabcolsep}{2pt}
 \begin{adjustbox}{max width=\linewidth}
 \begin{tabular}{@{}lcccccc@{}}
 \toprule
 Model & MAE $\downarrow$ ($\pm$\,half 95\% CI) & SSIM $\uparrow$ ($\pm$\,half 95\% CI) & PSNR $\uparrow$ ($\pm$\,half 95\% CI) & $R^2$ & Bias ($\pm$\,half 95\% CI) & LoA low -- LoA high \\
 \midrule
 \textbf{CAE3D (ours)} &  $\mathbf{0.066 \pm 0.001}$ & $\mathbf{0.80 \pm 0.01}$ & $\mathbf{24.0 \pm 0.1}$ & 0.35 & $-0.003 \pm 0.019$ & $-$0.115 -- 0.108 \\
 FNO\_3D &  \uline{$0.072 \pm 0.016$} & $0.78 \pm 0.06$ & $24.0 \pm 1.3$ & 0.47 & $-0.010 \pm 0.018$ & $-$0.116 -- 0.097 \\
 ResNet\_3D &  \uline{$0.072 \pm 0.008$} & \uline{$0.80 \pm 0.04$} & \uline{$23.2 \pm 0.8$} & 0.32 & $0.045 \pm 0.032$ & $-$0.119 -- 0.208 \\
 Patch\_3D & $0.078 \pm 0.016$ & $0.71 \pm 0.06$ & $22.9 \pm 1.4$ & 0.28 & $-0.037 \pm 0.019$ & $-$0.145 -- 0.070 \\
 CAE\_2D & $0.081 \pm 0.015$ & $0.68 \pm 0.09$ & $20.8 \pm 1.2$ & 0.58 & $0.020 \pm 0.023$ & $-$0.085 -- 0.125 \\
 Hybrid\_3D &  $0.120 \pm 0.013$ & $0.59 \pm 0.04$ & $18.8 \pm 0.7$ & $-$0.03 & $0.028 \pm 0.024$ & $-$0.113 -- 0.168 \\
 Cold\_3D &  $0.179 \pm 0.011$ & $0.37 \pm 0.02$ & $14.6 \pm 0.3$ & 0.31 & $-0.035 \pm 0.019$ & $-$0.143 -- 0.074 \\
 Residual\_3D &  $0.250 \pm 0.028$ & $0.31 \pm 0.02$ & $13.3 \pm 0.95$ & $-$7.44 & $-0.250 \pm 0.028$ & $-$0.407 -- $-$0.093 \\
 DDPM\_3D &  $0.749 \pm 0.027$ & $0.04 \pm 0.00$ & $1.12 \pm 0.11$ & $-$68.64 & $0.748 \pm 0.027$ & 0.592 -- 0.905 \\
SAM-Med3D\textsuperscript{$\dagger$} &  \textbf{$0.083 \pm 0.001$} & \textbf{$0.701 \pm 0.004$} & \textbf{$22.20 \pm 0.03$} & 0.45 & 0.015 & $-$0.099 -- 0.130 \\
 Med3DVLM\textsuperscript{$\dagger$} &  \uline{$0.098 \pm 0.003$} & \uline{$0.418 \pm 0.035$} & \uline{$15.98 \pm 1.86$} & 0.46 & $-$0.011 & $-$0.126 -- 0.105 \\
\bottomrule
 \end{tabular}
 \end{adjustbox}
 \end{table}

\paragraph{Seed-to-seed stability.}
Table~\ref{tab:fixed_test_three_seed} summarizes seed-to-seed variability for the primary trained-from-scratch 3D models on the fixed held-out test set. Unlike Table~\ref{tab:main_metrics}, which provides the main bootstrap-based model comparison across all methods, this table focuses on optimization stability across three independent training seeds. \textbf{Residual\_3D\_tips} (DDIM sampling, patch-based training, a cosine noise schedule) collapsed to a near-zero-residual solution across all three seeds despite genuinely distinct training; we report it for transparency only, not as evidence of stability (table caption; Appendix~C). \textbf{Patch\_3D}'s near-zero variance instead reflects a pretrained VAE shared across the three diffusion-stage seeds, not genuine seed-independence (table caption; Appendix~C).

\begin{table}[tb]
 \centering
 \footnotesize
 \caption[Three-seed stability on fixed held-out test set]{Seed-to-seed stability: cohort-mean MAE, SSIM, PSNR as mean$\pm$std over three seed-indexed runs for the primary trained-from-scratch 3D models (main comparison: Table~\ref{tab:main_metrics}); runs were independent except Patch\_3D, detailed below. \textsuperscript{$\ddagger$}Residual\_3D\_tips's three seeds are genuinely distinct checkpoints, but all collapsed to the same degenerate near-identity solution (predicted residual $\approx 0$); not a valid stability/performance result (\S\ref{sec:results_holdout}) and excluded from ranking. \textsuperscript{\S}For Patch\_3D, only the latent diffusion stage was seeded per run; the pretrained VAE was shared across all three fixed-test runs rather than retrained, so this is not a fully independent three-seed result---its near-zero SD reflects the shared VAE, not seed-independence ($K$-fold replication, Table~\ref{tab:kfold_rotating}, retrains the VAE per fold and shows nonzero variance).}
 \label{tab:fixed_test_three_seed}
 \begin{adjustbox}{max width=0.85\linewidth}
 \begin{tabular}{@{}lccc@{}}
 \toprule
 Model & MAE (mean $\pm$ std) $\downarrow$ & SSIM (mean $\pm$ std) $\uparrow$ & PSNR (mean $\pm$ std) $\uparrow$ \\
 \midrule
 \textbf{CAE3D (ours)} & $\mathbf{0.0663 \pm 0.0008}$ & $\mathbf{0.7986 \pm 0.0011}$ & $\mathbf{24.00 \pm 0.08}$ \\
 Residual\_3D\_tips\textsuperscript{$\ddagger$} & $0.0675 \pm 0.0000$ & $0.7885 \pm 0.0000$ & $23.98 \pm 0.00$ \\
 CAE3D-ES & $0.0722 \pm 0.0013$ & $0.7933 \pm 0.0009$ & $23.31 \pm 0.15$ \\
 Patch\_3D\textsuperscript{\S} & $0.0721 \pm 0.0000$ & $0.7235 \pm 0.0000$ & $23.32 \pm 0.00$ \\
 FNO\_3D & $0.0725 \pm 0.0001$ & $0.7744 \pm 0.0003$ & $23.87 \pm 0.02$ \\
 ResNet\_3D & $0.0767 \pm 0.0008$ & $0.7233 \pm 0.0056$ & $22.32 \pm 0.08$ \\
 Hybrid\_3D & $0.1038 \pm 0.0035$ & $0.6255 \pm 0.0160$ & $19.54 \pm 0.26$ \\
 Cold\_3D & $0.2476 \pm 0.0173$ & $0.2707 \pm 0.0390$ & $12.27 \pm 0.33$ \\
 \bottomrule
 \end{tabular}
 \end{adjustbox}
\end{table}

Among the trained-from-scratch models, \textbf{CAE3D (ours)} achieved the strongest overall held-out performance. Paired Wilcoxon tests with Holm adjustment (family of eight comparisons per metric, $N{=}32$ paired subjects) confirmed that CAE3D's MAE advantage was statistically significant over ResNet\_3D, Cold\_3D, DDPM\_3D, FNO\_3D, Hybrid\_3D, Patch\_3D, and Residual\_3D ($p_{\text{Holm}} < 0.05$), but not over the 2D middle-slice \textbf{CAE\_2D} baseline ($p_{\text{Holm}} = 0.080$); its SSIM and PSNR advantages were statistically significant over all eight comparators, including CAE\_2D. Full pairwise tables are in the supplementary material. Because CAE\_2D is a different-domain, middle-slice comparator (\S\ref{sec:metrics}), these significance tests are descriptive and supportive rather than evidence that volumetric modeling is statistically superior to slice-based modeling. Among the weaker baselines, \textbf{DDPM\_3D} failed catastrophically ($R^2=-68.64$, consistent with near-constant high-intensity output under the unconditioned diffusion schedule), and \textbf{Residual\_3D} collapsed toward the pre-ACZ image (bias $-0.250$), producing systematically underestimated post-ACZ intensities. \textbf{Cold\_3D} and \textbf{Hybrid\_3D} trained successfully but lagged the deterministic models. Foundation adapters (\textsuperscript{$\dagger$}) used a different evaluation protocol and are not directly comparable to in-house rows. For example, \textbf{SAM-Med3D} outperformed \textbf{Med3DVLM}, but neither matched the best deterministic results (Table~\ref{tab:main_metrics} caption; \S\ref{sec:foundation_baselines}). Regional summaries appear in Tables~\ref{tab:territory_delta_per_region} and \ref{tab:regional_deltacbf_vascular}.

\paragraph{Interpreting subject-level $R^2$.} CAE3D's per-subject $R^2$ was modest (0.35) despite its leading MAE, while \textbf{FNO\_3D} reached a higher $R^2$ (0.47) at a comparable MAE (0.072 vs.\ 0.066). This is not a contradiction: $R^2$ is computed on per-subject mean in-mask intensities (\S\ref{sec:metrics}) and reflects how well a model reproduces \emph{between-subject} variance in mean post-ACZ signal, whereas MAE and SSIM measure \emph{within-subject}, voxelwise fidelity. A model can reconstruct each subject's volume accurately while still explaining only a fraction of between-subject variance; plausible contributors include heterogeneous vasodilatory response magnitude, per-volume normalization effects, measurement variability, and post-ACZ information not identifiable from the baseline scan alone. We have not formally decomposed these contributions, nor analyzed the FNO\_3D/CAE3D gap as a bias-variance trade-off. We do not read $R^2=0.35$ as evidence that CAE3D recovers each patient's individualized CVR; together with the territory-level $\Delta$CBF compression discussed below, it indicates the model captures population-level structure more reliably than subject-specific vasodilatory variation, a distinction we return to in \S\ref{sec:discussion}.

\paragraph{Agreement and Bland-Altman.}
Bias and LoA appear per model in Table~\ref{tab:main_metrics}; Table~\ref{tab:bland_altman} (Appendix~C) additionally reports the SD underlying each LoA. Agreement broadly followed reconstruction quality: stable deterministic models showed near-neutral bias, while poorly performing diffusion models showed large bias and wide limits of agreement. CAE3D achieved the smallest absolute bias ($-$0.003), with LoA comparable to FNO\_3D and CAE\_2D ($\pm$0.11 for all three); bias should nonetheless be interpreted together with LoA and reconstruction error, since a small mean difference does not guarantee narrow subject-level spread.

\paragraph{Territory-level $\Delta$CBF (exploratory).}
Table~\ref{tab:territory_delta_per_region} reports cohort-mean $\Delta$CBF (\%) per territory for ground truth, CAE3D, and ResNet\_3D. CAE3D preserved territorial ordering with substantially smaller deviation from ground truth than the unstable diffusion baselines, while ResNet\_3D tended to overestimate vascular responses; consistent with deterministic autoencoders generally~\citep{goyal2026generating}, CAE3D compressed the predicted $\Delta$CBF dynamic range, most pronounced in the cerebellum and pons ($\sim$10-fold smaller than ground truth). In absolute normalized units ($\Delta_{\text{abs}}$), however, CAE3D retained a substantial fraction of the true territory-level change. Absolute $\Delta$CBF magnitudes should not be read as calibrated CVR values; conversion to physical units (ml/100\,g/min) requires inversion using the per-subject pre-ACZ CBF scale, which we have not performed in this study (\S\ref{sec:discussion_future}). Pooled vascular-mask means for nine 3D models (excluding 2D models, because the primary focus is full-volume 3D synthesis) appear in Table~\ref{tab:regional_deltacbf_vascular}.

\begin{table}[tb]
 \centering
 \footnotesize
 \caption[Per-territory $\Delta$CBF (fixed test)]{Cohort-mean $\Delta$CBF by vascular territory (fixed held-out, $N{=}32$). Percent $\Delta = (\bar{I}_{\text{post}}-\bar{I}_{\text{pre}})/\bar{I}_{\text{pre}}\times100$; absolute $\Delta_{\text{abs}} = \bar{I}_{\text{post}}-\bar{I}_{\text{pre}}$ (normalized $[0,1]$ units; both computed from territory-averaged intensities). CAE3D compresses percent $\Delta$ in high-dynamic-range territories (cerebellum, pons) but captures a substantial fraction of the true absolute change. ResNet\_3D overestimates absolute $\Delta$ in most territories.}
 \label{tab:territory_delta_per_region}
 \begin{adjustbox}{max width=\linewidth}
 \begin{tabular}{@{}lrrrrrr@{}}
 \toprule
 Territory & GT $\Delta$\% & \textbf{CAE3D} $\Delta$\% & ResNet $\Delta$\% & GT $\Delta_{\text{abs}}$ & \textbf{CAE3D} $\Delta_{\text{abs}}$ & ResNet $\Delta_{\text{abs}}$ \\
 \midrule
 left\_ACA         & 19.1 & 14.3 & 33.9 & 0.048 & 0.025 & 0.069 \\
 left\_MCA         & 32.4 & 18.4 & 40.4 & 0.051 & 0.033 & 0.057 \\
 left\_PCA         & 43.4 & 17.8 & 45.6 & 0.059 & 0.042 & 0.080 \\
 left\_cerebellum  & 187.6 & 18.8 & 80.8 & 0.060 & 0.051 & 0.087 \\
 left\_pons\_medulla & 115.2 & 19.2 & 76.6 & 0.023 & 0.041 & 0.040 \\
 right\_ACA        & 18.2 & 16.0 & 30.1 & 0.042 & 0.027 & 0.057 \\
 right\_MCA        & 23.6 & 17.6 & 37.2 & 0.040 & 0.030 & 0.057 \\
 right\_PCA        & 43.9 & 17.7 & 40.1 & 0.062 & 0.041 & 0.064 \\
 right\_cerebellum & 171.6 & 18.3 & 82.5 & 0.060 & 0.051 & 0.079 \\
 right\_pons\_medulla & 144.7 & 20.0 & 79.7 & 0.027 & 0.041 & 0.036 \\
 vascular\_territory & 25.8 & 16.6 & 33.9 & 0.046 & 0.030 & 0.059 \\
 \bottomrule
 \end{tabular}
 \end{adjustbox}
\end{table}

Supplementary Figure~S2 plots cohort-mean ground-truth territorial $\Delta$CBF for comparison.

 \begin{table}[tb]
 \centering
 \footnotesize
 \caption[Regional delta CBF]{Regional $\Delta$CBF (\%): cohort mean on the fixed held-out test set, vascular territory only. GT = ground truth; pred = model prediction.}
 \label{tab:regional_deltacbf_vascular}
 \begin{adjustbox}{max width=0.55\linewidth}
 \begin{tabular}{@{}lcc@{}}
 \toprule
 Model & mean $\Delta$CBF GT (\%) & mean $\Delta$CBF pred (\%) \\
 \midrule
 Cold\_3D & 25.8 & $-$77.9 \\
 DDPM\_3D & 25.8 & 350.7 \\
 FNO\_3D & 25.8 & 14.7 \\
 Hybrid\_3D & 25.8 & 38.3 \\
 Patch\_3D & 25.8 & $-$1.9 \\
 Residual\_3D & 25.8 & $-$99.8 \\
 ResNet\_3D & 25.8 & 33.9 \\
 \textbf{CAE3D (ours)} & 25.8 & 16.6 \\
 \bottomrule
 \end{tabular}
 \end{adjustbox}
 \end{table}

\footnotesize\noindent\textit{Note:} Extreme predicted $\Delta$CBF for DDPM\_3D and Residual\_3D reflects training instability under this setup (cf.\ ranking above).\par
\normalsize

\paragraph{Five-fold cross-validation.}
\label{sec:kfold_textbook}
To confirm that the fixed held-out ranking was not driven by a particular train/test split, we conducted a rotating $K{=}5$ cross-validation over the full cohort (train on $K{-}2$ folds, validate on one, test on one). Nine trained-from-scratch 3D models were evaluated under the same protocol; foundation adapters used a separate protocol. The fold-level cohort means in Table~\ref{tab:kfold_rotating} (Appendix~C) confirm CAE3D remained the top-performing model across partitions, though the ordering of the remaining models varied somewhat across folds.

\begin{figure}[tb]
    \centering
    \begin{minipage}{0.75\linewidth}
    \centering
    \ResolveSlideVis{\FProofFig}{\ProofFigResolved}%
    \ifx\ProofFigResolved\empty
      \fbox{\parbox{\linewidth}{\footnotesize\centering Figure asset unavailable.}}%
    \else
      \includegraphics[width=\linewidth]{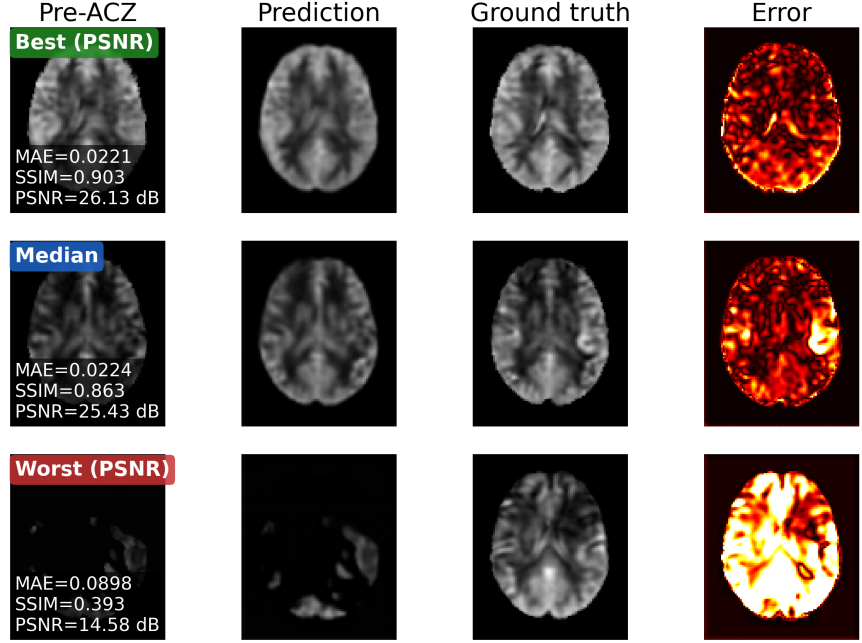}%
    \fi
    \end{minipage}
    \caption{\textbf{Qualitative reconstruction} (CAE3D, held-out test set): best, median, worst PSNR subjects (top to bottom). Columns: pre-ACZ, prediction, GT, error (middle axial slice). Per-subject MAE/SSIM/PSNR in row labels.}
    \label{fig:proof}
  \end{figure}

  \paragraph{Results on held-out test set.}
  \label{sec:results_holdout}
 Table~\ref{tab:regional_deltacbf_vascular} shows that CAE3D and FNO\_3D were closest to the ground-truth mean $\Delta$CBF among the evaluated 3D models. \textbf{Cold\_3D} underperformed the deterministic baselines on reconstruction fidelity (MAE/SSIM/PSNR) without collapsing as severely as \textbf{DDPM\_3D} or \textbf{Residual\_3D}, though its pooled-vascular $\Delta$CBF ($-$77.9\% vs.\ GT $+$25.8\%, Table~\ref{tab:regional_deltacbf_vascular}) shows comparably poor territory-level agreement. Supplementary Figures~S3--S4 show pre/post/change and a five-panel qualitative layout; the supplementary material also reports pipeline-unit MAE, a CAE3D-vs.-Cold\_3D permutation test ($p < 0.001$), regional Wilcoxon/FDR, RSM, and failure panels. High PSNR or favorable qualitative appearance (Figure~\ref{fig:proof}) does not by itself establish clinical appropriateness, because conventional reconstruction metrics may not capture errors that alter territory-level hemodynamic interpretation or treatment-related judgments.

  \begin{figure}[tb]
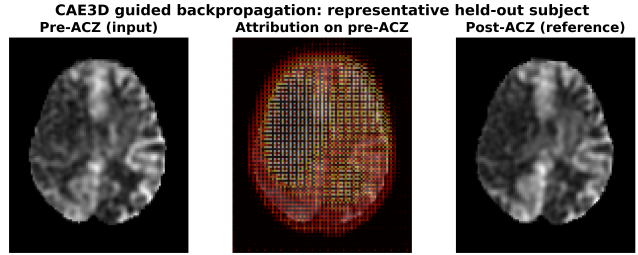

    \centering
    \begin{minipage}{0.55\linewidth}
      \centering
      \ResolveSlideVis{\FGuidePng}{\GuidePngResolved}%
      \ifx\GuidePngResolved\empty
        \ResolveSlideVis{\FGuidePdf}{\GuidePdfResolved}%
        \ifx\GuidePdfResolved\empty
          \fbox{\parbox{\linewidth}{\centering\small Figure asset unavailable.}}%
        \else
          \includegraphics[width=\linewidth]{\GuidePdfResolved}%
        \fi
      \else
        \includegraphics[width=\linewidth]{\GuidePngResolved}%
      \fi
    \end{minipage}
    \caption{\textbf{Guided backpropagation} (CAE3D, one test subject): pre-ACZ, attribution overlay (brighter = stronger influence), GT post-ACZ. Full guided-backpropagation maps are provided in the supplementary material.}
    \label{fig:guided_backprop}
  \end{figure}

 \section{Discussion}
 \label{sec:discussion}
 
\paragraph{Main findings and clinical interpretation.}
\label{sec:discussion_summary}
CAE3D synthesized post-ACZ perfusion maps from pre-ACZ input with reconstruction error that motivates further clinical validation, though not directly comparable to ASL scan-rescan repeatability (\S\ref{sec:results_holdout}). Deterministic conditional autoencoders outperformed diffusion-style variants and frozen-encoder adapters for this task. Qualitative inspection suggested prominent errors near mask boundaries and high-flow or low-signal regions, plausibly reflecting partial-volume, registration, and normalization effects not confirmed quantitatively via voxel-level error localization. Territory $\Delta$CBF remains a surrogate measure, not a validated CVR endpoint, complementing global metrics pending prospective work. CAE3D's modest per-subject $R^2$ (0.35, \S\ref{sec:results_holdout}), together with the territory-level $\Delta$CBF compression, indicates that CAE3D captures population-level structure in pre$\rightarrow$post mapping more reliably than subject-specific vasodilatory magnitude.
 
 \paragraph{Trustworthiness: uncertainty, agreement, and failure modes.}
 \label{sec:discussion_trustworthiness}
 This study reported seed variability, bootstrap CIs, and Bland-Altman bias/LoA, since correlation alone does not imply acceptable bias. Anticipated failure modes include registration/atlas mismatch, unstable $\Delta$CBF at low baseline signal, mask-boundary artifacts, and scanner or cohort shift; future evaluation should pair outputs with predefined quality checks and human review.
 
 \paragraph{Transparency, explainability, and auditing.}
 \label{sec:discussion_explainability}
Because CAE3D may influence downstream CVR interpretation, we provided two explainability analyses: guided backpropagation for one held-out subject (Figure~\ref{fig:guided_backprop}), and representational similarity analysis (RSM) of CAE3D bottleneck features across the 32 test subjects (supplementary material). Alignment with clinical subgroups was not tested because laterality and severity metadata were unavailable; curated good/typical/failure cases are more clinically informative than best-case examples. \textbf{On ethics and responsible use}, synthesized maps are research outputs for retrospective investigation, not decision-support tools or substitutes for clinical imaging, until prospective validation establishes their intended role. Data sharing must follow PHI/de-identification and drift-monitoring governance policies.
 
 \paragraph{Limitations.}
 \label{sec:discussion_limitations}
This study has several limitations. This is internal validation, not deployment-ready evidence: results may not generalize across scanners, ASL sequences, or sites, since we did not stress-test out-of-distribution or external cohorts; affine registration and atlas readouts can also degrade when anatomy departs from template space. $[0,1]$ normalization aids training but yields unitless MAE; inversion to physical CBF scales for clinical reporting has not been performed in this study (\S\ref{sec:discussion_future}). Region-level FDR was limited to one model pair; we did not run voxel/cluster permutation tests, confounder-adjusted analyses, calibrated uncertainty maps, or prospective workflow validation. The territory-level $\Delta$CBF compression noted in \S\ref{sec:results_holdout} may partly reflect the L1+SSIM training objective, which prioritizes average reconstruction fidelity over preservation of extreme vasodilatory responses; territories with inherently high GT $\Delta$CBF (cerebellum, pons/medulla) show the largest percent-scale compression. Finally, because the study cohort enrolled only patients who completed the two-scan protocol, no patient for whom ACZ was contraindicated or withheld under the study protocol is represented in the training or test data. Therefore, the prevalence of ACZ contraindications in the broader Moyamoya population and model performance in that population remain unknown. An independent surrogate for vasodilatory response (e.g., BOLD CVR from a hypercapnic/breath-hold challenge, or a deep-learning-based drug-free CVR estimate~\citep{ChenPredictingCVRRadiology2020}) would extend evaluation to contraindicated patients without administering ACZ (planned future work). ASL quality is limited by low signal-to-noise ratio and sensitivity to motion, scanner calibration, and acquisition parameters; a systematic study of these effects on synthesis fidelity is planned future work. No consensus CVR threshold for Moyamoya bypass surgery exists~\citep{SeeStoutActaNeurochir2023}; decisions integrate imaging, anatomy, symptoms, and other factors rather than a single threshold~\citep{NguyenFrontSurg2022}. Consequently, this study establishes synthesis feasibility, not surgical decision support or replacement of the ACZ challenge.
 
 \paragraph{Future directions.}
 \label{sec:discussion_future}
Immediate next steps include multicenter external validation and prospective, blinded clinician-in-the-loop evaluation. Broader directions include physical-unit CBF error reporting after inversion, range-aware training for the compression noted above, and full foundation-model fine-tuning. Territory-aware models may improve regional calibration at the cost of less training data per territory and added inference complexity. Flow matching~\citep{LipmanFlowMatching2023} is a relevant alternative given DDPM-family instability. Baseline-to-challenge synthesis warrants separate evaluation in intracranial atherosclerosis and sickle cell cerebrovascular disease.

 \bibliography{sample}

 \section*{Appendix A. Reproducibility}
 
 Preprocessing, training, and evaluation were implemented in Python using PyTorch and MONAI~\citep{cardoso2022monai}.
 In-house encoder-decoder and diffusion denoisers use MONAI's \texttt{UNet} / \texttt{DiffusionModelUNet} with task-specific configuration and recipes as in \S\ref{sec:baselines} (architecture details: \S\ref{sec:proposed}).
 The train/validation/test split is fixed (random seed 42) for all models. As described in \S\ref{sec:methods}, Table~\ref{tab:main_metrics} and Table~\ref{tab:bland_altman} report the single-instance evaluation for each in-house model, distinct from the three-seed evaluation in Table~\ref{tab:fixed_test_three_seed}. For every in-house model, seed 42 was predesignated (not selected post hoc from among the three training seeds) as the primary instance; within that seed, the checkpoint with the highest validation PSNR was retained, and the test set was not used in checkpoint selection. Table~\ref{tab:primary_checkpoints} lists this rule per model. For \textbf{CAE3D}, $R^2$, Bland-Altman bias, and LoA in Tables~\ref{tab:main_metrics} and~\ref{tab:bland_altman} were verified by direct re-evaluation of the seed-42 checkpoint. The code repository at \url{https://github.com/TheClassicTechno/cae3d-moyamoya-cbf-synthesis} documents the exact checkpoint file and training configuration for \textbf{CAE3D} (no early stopping), \textbf{CAE3D-ES} (early stopping), and every other baseline.

\begin{table}[H]
\centering
\footnotesize
\caption[Primary-checkpoint selection rule per model]{Primary-checkpoint selection rule for each in-house model's rows in Tables~\ref{tab:main_metrics} and~\ref{tab:bland_altman}. Seed 42 was predesignated for every model before training (not chosen post hoc from among the three seeds); within that seed, the checkpoint with the highest validation PSNR was retained, with no test-set involvement in selection. Exact checkpoint filenames and configurations are documented with the code release.}
\label{tab:primary_checkpoints}
\begin{tabular}{@{}lll@{}}
\toprule
Model & Primary seed & Checkpoint-selection rule \\
\midrule
CAE3D & 42 (predesignated) & Highest validation PSNR, no early stopping \\
CAE3D-ES & 42 (predesignated) & Highest validation PSNR, early stopping (10-epoch plateau) \\
ResNet\_3D & 42 (predesignated) & Highest validation PSNR \\
CAE\_2D & 42 (predesignated) & Highest validation PSNR \\
FNO\_3D & 42 (predesignated) & Highest validation PSNR \\
Patch\_3D & 42 (predesignated) & Highest validation PSNR \\
Hybrid\_3D & 42 (predesignated) & Highest validation PSNR \\
Cold\_3D & 42 (predesignated) & Highest validation PSNR \\
Residual\_3D & 42 (predesignated) & Highest validation PSNR \\
DDPM\_3D & 42 (predesignated) & Highest validation PSNR \\
\bottomrule
\end{tabular}
\end{table}

\begin{algorithm}[H]
\caption{CAE3D training and inference procedure}
\label{alg:cae3d}
\KwIn{paired volumes $(x_{\mathrm{pre}}, x_{\mathrm{post}})$}
\KwOut{predicted post-ACZ volume $\hat{x}_{\mathrm{post}}$}
Preprocess all pairs: affine registration to MNI, brain masking, normalization to $[0,1]$, and pad/crop to a valid grid\;
\For{epoch $=1$ to $50$}{
  $\hat{x}_{\mathrm{post}} \leftarrow \mathrm{CAE3D}(x_{\mathrm{pre}})$\;
  $\mathcal{L} \leftarrow \|\hat{x}_{\mathrm{post}}-x_{\mathrm{post}}\|_{1} + \bigl(1-\mathrm{SSIM}(\hat{x}_{\mathrm{post}},x_{\mathrm{post}})\bigr)$\;
  update parameters with Adam ($10^{-3}$)\;
  update the retained checkpoint if validation PSNR improves\;
}
Inference: load best checkpoint and compute $\hat{x}_{\mathrm{post}}=\mathrm{CAE3D}(x_{\mathrm{pre}})$\;
Evaluate full-brain MAE/SSIM/PSNR, agreement (Bland-Altman), and regional $\Delta$CBF summaries\;
\end{algorithm}

 Supplementary material includes:
\begin{itemize}[noitemsep,topsep=2pt,parsep=0pt,leftmargin=1.2em]
  \item Holm-adjusted pairwise Wilcoxon tables (MAE/SSIM/PSNR)
  \item TIPS residual-diffusion note
  \item Full guided-backprop maps (one panel shown in Figure~\ref{fig:guided_backprop})
  \item RSM heatmap and table; loss ablation; pipeline-unit MAE
  \item Permutation and regional FDR analyses; failure-case panels
  \item \textbf{Per-territory MAE/SSIM/PSNR table for CAE3D} (mean$\pm$std per atlas territory; Table~\ref{tab:territory_mae_ssim_psnr}, Appendix~C)
  \item Inputs underlying the five-fold summary (Table~\ref{tab:kfold_rotating})
  \item Foundation-baseline aggregates (Appendix~C); per-territory $\Delta$CBF rows for Residual\_3D\_tips
\end{itemize}
Training defaults: Adam $10^{-3}$, batch size~2, 50~epochs, checkpoint by validation PSNR. Bootstrap resamples $B{=}2000$. 

Preprocessing: \S\ref{sec:data}. Regeneration scripts and file-level provenance are documented in the code repository above.

 \section*{Appendix B. Supplementary figures}

 Regional and extra qualitative figures match the slide-visual assets used for the main-text figures. Figure~S1 below shows binary mask unions on one middle axial slice, using the same slice convention as Figure~\ref{fig:vascular_territories_colormap} in the main text.

 \begin{figure}[H]
   \centering
   \includegraphicsSlideVis[width=\linewidth]{\FSlideRegional}
   \par\vspace{2pt}
   {\footnotesize\textbf{S1:} Atlas mask unions (left hem., right hem., pooled vascular; middle axial slice).}
 \end{figure}

Figure~S2 next plots cohort-mean ground-truth territorial $\Delta$CBF (\%) for comparison with Table~\ref{tab:territory_delta_per_region} in the main text.

 \begin{figure}[H]
   \centering
   \includegraphicsSlideVis[width=\linewidth]{\FSlideGtDelta}
   \par\vspace{2pt}
   {\footnotesize\textbf{S2:} Cohort mean ground-truth territorial $\Delta$CBF (\%) (one bar per summary region).}
 \end{figure}

Figures~S3 and S4 turn from these territory-level summaries to per-subject qualitative examples. The paired pre-ACZ, post-ACZ, and voxelwise post-minus-pre change maps for this example subject are shown in Figure~\ref{fig:image_context} in the main text; that figure is not reproduced here.

Figure~S4 extends this to a five-panel qualitative layout (pre, post, prediction, mask, error) for one subject, complementing the best/median/worst PSNR panels in Figure~\ref{fig:proof} of the main text.

 \begin{figure}[H]
   \centering
   \includegraphicsSlideVis[width=\linewidth]{\FSlideFive}
   \par\vspace{2pt}
   {\footnotesize\textbf{S4:} Five-panel qualitative summary (pre, post, prediction, mask, error).}
 \end{figure}

\section*{Appendix C. $K$-fold sources and foundation baselines}

\paragraph{Notes for Table~\ref{tab:main_metrics} (main text).}
Rows marked \textsuperscript{$\dagger$} are frozen-encoder foundation adapters. For these rows, MAE/SSIM/PSNR are reported as mean$\pm$std over training seeds (eight for Med3DVLM; three for SAM-Med3D), and bold/underline on those three columns compares only the two adapters. Their $R^2$ and Bland--Altman bias/LoA are point estimates computed from the seed-42 checkpoint only (sources in Table~\ref{tab:foundation_supplement}). Bootstrap bias CIs were not computed for these rows. For PSNR, Table~\ref{tab:main_metrics} bolds only CAE3D: FNO\_3D matches the cohort mean to one decimal place (24.0\,dB) but is lower in the three-seed summary reported in Table~\ref{tab:fixed_test_three_seed}.

\paragraph{Bland-Altman detail (Std of paired differences).}
Table~\ref{tab:bland_altman} reproduces the per-model Bias and LoA already given in Table~\ref{tab:main_metrics} (main text) alongside the SD of paired differences underlying each LoA ($\mathrm{LoA}=\mathrm{Bias}\pm1.96\times\mathrm{SD}$).

 \begin{table}[H]
 \centering
 \caption[Bland-Altman agreement]{Bland-Altman agreement (predicted $-$ ground truth, normalized $[0,1]$ scale): mean difference (Bias $\rightarrow 0$), std of differences (Std $\downarrow$), and 95\% limits of agreement (LoA $= \pm 1.96 \times$ SD). \textbf{Bold}/\underline{underline}: smallest/second-smallest $|$Bias$|$.}
 \label{tab:bland_altman}
 \footnotesize
 \setlength{\tabcolsep}{2pt}
 \begin{adjustbox}{max width=0.72\linewidth}
 \begin{tabular}{@{}lcccc@{}}
 \toprule
 Model & Bias $\rightarrow 0$ & Std $\downarrow$ & LoA$_{\text{low}}$ & LoA$_{\text{high}}$ \\
 \midrule
 Cold\_3D & $-$0.035 & 0.055 & $-$0.143 & 0.074 \\
 DDPM\_3D & 0.748 & 0.080 & 0.592 & 0.905 \\
 \uline{FNO\_3D} & \uline{$-$0.010} & \textbf{0.054} & $-$0.116 & 0.097 \\
 Hybrid\_3D & 0.028 & 0.072 & $-$0.113 & 0.168 \\
 Patch\_3D & $-$0.037 & 0.055 & $-$0.145 & 0.070 \\
 Residual\_3D & $-$0.250 & 0.080 & $-$0.407 & $-$0.093 \\
 ResNet\_3D & 0.045 & 0.083 & $-$0.119 & 0.208 \\
 CAE\_2D & 0.020 & 0.054 & $-$0.085 & 0.125 \\
 \textbf{CAE3D (ours)} & $\mathbf{-0.003}$ & 0.057 & $-$0.115 & 0.108 \\
 \bottomrule
 \end{tabular}
 \end{adjustbox}
 \end{table}

\paragraph{$K$-fold replication.}
Table~\ref{tab:kfold_rotating} summarizes fold-level cohort means: each model’s five test folds yield one cohort mean per metric, and the table reports mean$\pm$std across those five values. Fold-level outputs are stored with the released training scripts. The same protocol can be reproduced from the code release, with optional checks of per-fold tables before exporting LaTeX rows.

\begin{table}[H]
 \centering
 \footnotesize
 \caption[Rotating $K$-fold summary ($K{=}5$)]{Five-fold rotation over the full cohort: entries are mean$\pm$std of the five fold-level cohort means. \textbf{Bold}/underline: best/second-best. \textsuperscript{$\ddagger$}Residual\_3D\_tips collapsed to a near-identity (predicted residual $\approx 0$) solution on this cohort (\S\ref{sec:results_holdout}); its row reflects that collapse, evaluated across five different test folds, and is excluded from best/second-best ranking. The collapse traces to a numerical-stability issue in the shared DDIM/cosine-schedule sampling code that is checkpoint-independent (\S\ref{sec:tips_collapse_verification}); we directly verified this mechanism against a trivial zero-residual baseline for the fixed-test three-seed instance (Table~\ref{tab:fixed_test_three_seed}) but did not independently re-verify each of the five $K$-fold instances against that baseline.}
 \label{tab:kfold_rotating}
 \setlength{\tabcolsep}{3pt}
 \begin{adjustbox}{max width=\linewidth}
 \begin{tabular}{@{}lccc@{}}
 \toprule
 Model & MAE $\downarrow$ & SSIM $\uparrow$ & PSNR $\uparrow$ (dB) \\
 \midrule
 \textbf{CAE3D (ours)} & \textbf{$0.0746 \pm 0.0049$} & \textbf{$0.7943 \pm 0.0115$} & \textbf{$22.78 \pm 0.42$} \\
 CAE3D-ES & \uline{$0.0774 \pm 0.0056$} & \uline{$0.7899 \pm 0.0124$} & $22.42 \pm 0.43$ \\
 Residual\_3D\_tips\textsuperscript{$\ddagger$} & $0.0804 \pm 0.0046$ & $0.7774 \pm 0.0097$ & $22.74 \pm 1.23$ \\
 ResNet\_3D & $0.0855 \pm 0.0030$ & $0.7237 \pm 0.0141$ & $21.20 \pm 0.22$ \\
 FNO\_3D & $0.0856 \pm 0.0047$ & $0.7601 \pm 0.0100$ & $22.27 \pm 0.32$ \\
 Patch\_3D & $0.0856 \pm 0.0042$ & $0.7161 \pm 0.0122$ & $21.69 \pm 0.33$ \\
 Hybrid\_3D & $0.1070 \pm 0.0041$ & $0.6295 \pm 0.0058$ & $19.12 \pm 0.24$ \\
 Cold\_3D & $0.2506 \pm 0.0431$ & $0.2739 \pm 0.0409$ & $12.15 \pm 1.42$ \\
 DDPM\_3D & $0.6150 \pm 0.0527$ & $0.0116 \pm 0.0032$ & $2.54 \pm 0.42$ \\
 \bottomrule
 \end{tabular}
 \end{adjustbox}
\end{table}

\paragraph{Foundation adapter aggregates.}
Table~\ref{tab:foundation_supplement} reports mean$\pm$std MAE/SSIM/PSNR for frozen-encoder adapters and is intended to be read alongside the foundation rows in Table~\ref{tab:main_metrics}.
 
 \begin{table}[H]
  \centering
  \footnotesize
\caption{Foundation-style baselines on the combined cohort split. Med3DVLM and SAM-Med3D (frozen image encoder + trained 3D decoder) use pre$\rightarrow$post ASL perfusion maps with the same full-brain evaluator as \textbf{CAE3D}; mean$\pm$std is over training seeds (eight for Med3DVLM; three for SAM-Med3D). Full filenames and seed lists are documented with the code release.}
  \label{tab:foundation_supplement}
  \begin{adjustbox}{max width=\linewidth}
  \begin{tabular}{@{}p{2.6cm}p{3.4cm}ccc@{}}
  \toprule
  Model & What is scored & MAE & SSIM & PSNR (dB) \\
  \midrule
 Med3DVLM (DCFormer+dec.) & Pred.\ vs.\ GT post-ACZ ASL perfusion maps (full-brain) & $0.098 \pm 0.003$ & $0.418 \pm 0.035$ & $15.98 \pm 1.86$ \\
 SAM-Med3D (enc.\ froz.\ + dec.) & Pred.\ vs.\ GT post-ACZ ASL perfusion maps (full-brain) & $0.083 \pm 0.001$ & $0.701 \pm 0.004$ & $22.20 \pm 0.03$ \\
  \bottomrule
  \end{tabular}
  \end{adjustbox}
 \end{table}

 \paragraph{Per-territory MAE, SSIM, PSNR for CAE3D.}
Table~\ref{tab:territory_mae_ssim_psnr} reports cohort mean$\pm$std per atlas territory for \textbf{CAE3D} on the held-out test set ($N{=}32$). Territory names follow the atlas of \citet{LiuVascularAtlasScientificData} registered to MNI 2\,mm.

\begin{table}[H]
\centering
\footnotesize
\caption[Per-territory MAE/SSIM/PSNR for CAE3D]{CAE3D per-territory reconstruction metrics (held-out test set, $N{=}32$): cohort mean$\pm$std over subjects within each atlas territory. Metrics computed full-brain within each atlas mask on $[0,1]$ normalized data.}
\label{tab:territory_mae_ssim_psnr}
\begin{adjustbox}{max width=\linewidth}
\begin{tabular}{@{}lccc@{}}
\toprule
Territory & MAE (mean$\pm$std) $\downarrow$ & SSIM (mean$\pm$std) $\uparrow$ & PSNR (mean$\pm$std) $\uparrow$ \\
\midrule
left ACA         & $0.0827 \pm 0.0489$ & $0.9879 \pm 0.0120$ & $20.79 \pm 4.14$ \\
left MCA         & $0.0891 \pm 0.0571$ & $0.9763 \pm 0.0248$ & $20.22 \pm 4.17$ \\
left PCA         & $0.0939 \pm 0.0856$ & $0.9908 \pm 0.0142$ & $20.78 \pm 4.74$ \\
left cerebellum  & $0.1003 \pm 0.1062$ & $0.9943 \pm 0.0085$ & $21.02 \pm 5.27$ \\
left pons/medulla& $0.0866 \pm 0.0792$ & $0.9982 \pm 0.0031$ & $21.82 \pm 5.00$ \\
right ACA        & $0.0833 \pm 0.0570$ & $0.9879 \pm 0.0124$ & $20.87 \pm 4.23$ \\
right MCA        & $0.0893 \pm 0.0588$ & $0.9755 \pm 0.0233$ & $20.24 \pm 4.04$ \\
right PCA        & $0.0973 \pm 0.0881$ & $0.9911 \pm 0.0133$ & $20.52 \pm 4.73$ \\
right cerebellum & $0.1022 \pm 0.1058$ & $0.9944 \pm 0.0082$ & $20.83 \pm 5.41$ \\
right pons/medulla& $0.0930 \pm 0.0875$ & $0.9982 \pm 0.0029$ & $21.31 \pm 5.03$ \\
vascular territory& $0.0897 \pm 0.0606$ & $0.8910 \pm 0.0879$ & $19.96 \pm 3.81$ \\
\bottomrule
\end{tabular}
\end{adjustbox}
\end{table}

 Three-seed held-out and $K$-fold metrics for \textbf{Residual\_3D\_tips} are reported in Tables~\ref{tab:fixed_test_three_seed} and \ref{tab:kfold_rotating}, respectively, with the collapse caveat noted in both captions; it is not included in the territory-level $\Delta$CBF comparison (Table~\ref{tab:regional_deltacbf_vascular}), which uses \textbf{Residual\_3D} instead.

\paragraph{Residual\_3D\_tips collapse: checkpoint-level verification.}
\label{sec:tips_collapse_verification}
We confirmed via direct re-evaluation of each of the three saved seed-42/123/456 checkpoints that they are genuinely distinct trained weights (distinct checkpoint files, distinct training-loss trajectories), ruling out a simple checkpoint-reuse artifact. Despite this, all three converged to a degenerate solution in which the predicted residual is driven to approximately zero, so the reconstructed output reduces to the clipped pre-ACZ input regardless of checkpoint; independently computing MAE/SSIM/PSNR for a trivial ``predict zero residual'' baseline on this cohort reproduces the reported Table~\ref{tab:fixed_test_three_seed} numbers to within floating-point precision. This is the same collapse failure mode documented for \textbf{Residual\_3D} (\S\ref{sec:results_holdout}); the TIPS stabilization measures did not prevent it here. The proximate mechanism is a numerical-stability issue in the cosine noise-schedule construction shared by every TIPS checkpoint: the schedule's cumulative-product term can go slightly negative near the tail (a floating-point artifact), producing NaN values that are floored to zero at each DDIM sampling step, driving the predicted residual toward zero regardless of the underlying trained weights. Because this is a property of the shared sampling code rather than of any individual trained instance, it is expected to affect the $K$-fold instances (Table~\ref{tab:kfold_rotating}) by the same mechanism; we verified this directly against a trivial zero-residual baseline only for the fixed-test three-seed instance above, not independently for each $K$-fold checkpoint.

\paragraph{Patch\_3D near-zero fixed-test variance: implementation detail.} Patch\_3D's architecture couples a VAE (encoding overlapping patches into a latent space) with a latent diffusion stage (\S\ref{sec:baselines}). For the fixed-test three-seed experiment, the same pretrained VAE was reused across all three diffusion-stage seeds; only the latent diffusion component varied by seed. Because the shared VAE reconstruction dominates the decoded full-volume output at this patch/stride configuration, genuine per-seed differences confined to the diffusion stage are not resolved at the precision reported in Table~\ref{tab:fixed_test_three_seed}. This is specific to the fixed-test run rather than the architecture: the $K$-fold experiment (Table~\ref{tab:kfold_rotating}) retrained a fold-specific VAE for each fold and shows the expected nonzero seed-to-seed variance for Patch\_3D.

 \acks{We thank Prof. Kilian Pohl and Prof. Ehsan Adeli of Stanford University for helpful guidance and discussion during this project. This work is supported by American Heart Association Career Development Award \#24CDA1266771.}

\section*{LLM Use Disclosure}
Portions of this manuscript were drafted and revised with the assistance
of an AI language model, Claude (Anthropic), consistent with the MLHC LLM
Use Policy. AI assistance included drafting and editing prose, and
assistance with code used for model implementation, training/evaluation
scripts, and figure generation. All experimental results, figures, and
scientific claims were reviewed and validated by the authors, who take
full responsibility for the accuracy of all content in this manuscript.

 \end{document}